\documentclass[psfig]{aa}
\input psfig.tex

\newfont{\sfsl}{cmssqi8 scaled 1200}
\newcommand{\gcs}{{\sfsl HIFLUGCS}}

\begin{document}                                                                

\twocolumn

   \title{A Systematic Study of X-Ray Substructure of Galaxy Clusters
	Detected in the RO\-SAT All-Sky Survey }

   \titlerunning{A Systematic Study of X-Ray Substructure\thanks{This
  research has made use of the RO\-SAT All-Sky Survey Data which have
  been processed at MPE}}

   \author{P.\,Schuecker\,$^{(1)}$, 
           H.\,B\"ohringer\,$^{(1)}$, 
           T.H.\,Reiprich\,$^{(1)}$ and 
           L.\,Feretti\,$^{(2)}$}     

   \authorrunning{Schuecker et al.}

   \offprints{Peter Schuecker\\ peters@mpe.mpg.de} 

   \institute{
    $^{(1)}$ Max-Planck-Institut f\"ur extraterrestrische Physik,
             Garching, Germany\\
    $^{(2)}$ Istituto di Radioastronomia CNR, Bologna, Italy.}

   \date{Received ..... ; accepted .....}                            

   \markboth {A Systematic Study of X-Ray Substructure}{}

\abstract{Results of a systematic study of substructure in X-ray
surface brightness distributions of a combined sample of 470
REFLEX$+$BCS clusters of galaxies are presented. The fully automized
morphology analysis is based on data of the 3rd processing of the
RO\-SAT All-Sky survey (RASS-3). After correction for several
systematic effects, $52\pm 7$ percent of the REFLEX$+$BCS clusters are
found to be substructured in metric apertures of 1\,Mpc radius
($H_0=\,50\,{\rm km}\,{\rm s}^{-1}\,{\rm Mpc}^{-1}$). Future
simulations will show statistically which mass spectrum of major and
minor mergers contributes to this number. Another important result is
the discovery of a substructure-density relation, analogous to the
morphology-density relation for galaxies. Here, clusters with
asymmetric or multi-modal X-ray surface brightness distributions are
located preferentially in regions with higher cluster number
densities.  The substructure analyses techniques are used to compare
the X-ray morphology of 53 clusters with radio halos and relics, and
22 cooling flow clusters with the REFLEX$+$BCS reference sample. After
careful equalization of the different `sensitivities' of the
subsamples to substructure detection it is found that the halo and
relic sample tends to show more often multi-modal and elongated X-ray
surface brightness distributions compared to the REFLEX$+$BCS
reference sample. The cooling flow clusters show more often circular
symmetric and unimodal distributions compared to the REFLEX$+$BCS and
the halo/relic reference samples. Both findings further support the
idea that radio halos and relics are triggered by merger events, and
that pre-existing cooling flows might be disrupted by recent major
mergers.   }

\maketitle

\section{Introduction}\label{INTRO}

Current structure formation scenarios suggest an hierarchical growth
of cosmic objects (e.g., Peebles 1980, White 1996) where the merging
of subclumps turns out to be a fundamental process. Linear theory
predicts that clustering in critical density universes continues to
grow up to present-day redshifts, and structure formation begins to
decline already at $z\approx \Omega_0^{-1}-1$ in low density
universes, where $\Omega_0$ is the present value of the normalized
cosmic matter density. Compared to the Einstein-de Sitter case,
clusters in low-$\Omega_0$ universes are thus expected to be more
relaxed, less substructured, and less elongated as shown by the
simulations of the Virgo Consortium (see Thomas et al. 1998) and the
simulations of Evrard et al. (1993), Crone, Evrard \& Richstone
(1996), Mohr et al. (1995), etc.  The frequencies of subclumps
(substructures) and elongations are thus useful statistical quantities
with a direct relation to cosmology. See also the analytic work of,
e.g., Richstone, Loeb \& Turner (1992) and Lacey \& Cole (1993). Note,
however, that the effects are difficult to measure so that the
resulting constraints on structure formation models are presently less
stringent compared to, e.g., a direct measurement of the power
spectrum of cluster number density fluctuations as presented in, e.g.,
Schuecker et al. (2001).

Cluster mergers produce moderately supersonic shocks, compressing and
heating the intracluster gas, and increasing pressure and
entropy. This can be measured as local distortions of the spatial
distribution of X-ray temperature and surface brightness (e.g.,
Schindler \& M\"uller 1993, Roettiger et al. 1997). Moreover, mergers
affect cluster X-ray luminosity, magnetic field, and electron/ion
non-equipartition (see, e.g., Schindler \& M\"uller 1993, Roettiger,
Stone \& Burns 1999, Takizawa 2000, Ricker \& Sarazin
2001). Therefore, studies of individual clusters provide a wealth of
new and useful information on the physics of the merger process, e.g.,
from spatially resolved X-ray spectroscopy. 

A more detailed study would thus ideally use deep pointed X-ray
observations to get high signal-to-noise temperature and surface
brightness maps. Unfortunately, X-ray data from public archives
generally allow precise measurements on interesting but not
necessarily representative clusters of galaxies. For the majority of
the clusters needed for a statistically representative sample,
detailed studies are at present and in the near future not
achievable. What is needed for cosmological investigations is a
systematic and comprehensive study of substructure characteristics for
a large set of clusters compiled in an homogeneous way.

Concerning the projected X-ray surface brightness distribution, merger
events cause multiple X-ray luminosity peaks, isophote twisting with
centroid shifts, elongations, and other irregularities. One of the
first systematic studies of such cluster X-ray morphologies was
undertaken by Jones \& Forman (1999) using the spatial surface
brightness distributions of targeted and serendipitous clusters
obtained with the {\it Einstein} imaging proportional counter
(IPC). Their visual classifications of the X-ray iso-intensity
contours of 208 clusters with $z\le 0.15$, are useful, but
subjective. For refined statistical analyses it should be supplemented
by a more homogeneous sample selection and by a more objective, i.e.,
quantitative method to analyse the cluster morphology. Along this
line, Mohr et al. (1995) measured emission-weighted variations of the
cluster centroid for substructure detection, but only for a
comparatively small sample of 65 bright {\it Einstein} clusters. The
remaining studies not mentioned here are restricted to even smaller
cluster samples.

One possibility towards a systematic study of substructure is offered
by the large area RO\-SAT All-Sky Survey (RASS, Tr\"umper 1993, Voges
et al. 1999). However, both the energy resolution of 40 percent FW\-HM
at 1\,keV and the limited energy range $0.1\le E \le 2.4\,{\rm keV}$
of the RO\-SAT position-sensitive proportional counter (PSPC) limit
temperature measurements of the intracluster gas, especially for
$kT\ge 4\,{\rm keV}$.  This and the fact that the small number of
RASS-3 photons counted in the direction of a typical galaxy cluster in
our samples do neither allow the determination of useful temperature
maps nor the proper quantification of the amount of substructure,
e.g., in the form of mass estimates of the individual subclumps for
the majority of a fair cluster sample (but see Sect.\,\ref{DISCUSS}).
The present analysis thus gives more weight to a representative study
of a large sample of clusters, mainly restricted to substructure
detection and significance determination in terms of significance for
the degree of substructure for a whole sample or for specific
subsamples.

In a series of papers we want to study the morphology of a large set
of X-ray clusters in a systematic way. The present paper investigates
the morphology of galaxy clusters using the X-ray surface brightness
distributions extracted from data obtained in the course of the 3rd
processing of the RO\-SAT All-Sky Survey (RASS-3). Further papers will
concentrate on the study of alignment effects of the major cluster
axes, and on detailed comparisons with large scale, high resolution
numerical simulations. The morphological analyses are mainly
restricted to galaxy clusters of the two largest and almost complete
X-ray cluster surveys finished to date: the RO\-SAT-ESO Flux-Limited
X-ray (REFLEX) cluster survey (B\"ohringer et al. 2001) in the South,
and a survey yielding the Brightest Cluster Sample (BCS, Ebeling et
al. 1998) in the North.

In the present investigation we also describe more detailed studies of
specific cluster types with so-called radio halos and radio relics,
and signatures of cooling flows. Reviews concerning the latter cluster
type can be found in, e.g., Fabian, Nulsen \& Canizares (1984), and
Fabian (1994), but see also Makishima et al. (2001). A few comments
concerning the former two cluster types will be given in the following
(see also the conference proceedings edited by B\"ohringer, Feretti \&
Schuecker 1999).

Willson (1970) discovered diffuse, cluster-wide, steep spectrum
synchrotron emission associated with the intra\-cluster plasma
medium. Recent lists of these objects can be found in, e.g., Feretti
(1999), Giovannini (1999), Giovannini, Tordi \& Feretti (1999), Owen,
Morrison \& Voges (1999), and Giovannini \& Feretti (2000). The radio
morphology is classified into radio halos and relics depending on
whether (almost unpolarized) radio emission is detected throughout the
cluster, or (higher polarized) radio emission is detected in the
cluster periphery. Radio halos and relics are predominantly found in
rich compact clusters with Bautz-Morgan types II-III (Hanisch 1982),
have high X-ray luminosities and temperatures, in most cases no
cooling flow signature, and large cluster X-ray core radii. It is
suggested (see, e.g., Harris et al. 1980, Burns et al. 1995, Feretti
\& Giovannini 1996, and the review of Sarazin 2001) that radio halos
form during the merging of subclusters, accelerating (existing
relativistic) particles in shocks formed in the intracluster gas,
although additional processes seem to be necessary. Because of the
comparatively small sample sizes involved (see Sect.\,\ref{SAMPLES})
we refer in the following only to the combined radio halo/relic
sample, neglecting the interesting differences between the two cluster
types. 

Section\,\ref{METHODS} describes the methods used to detect
substructure in two-dimensional photon distributions and to assign
statistical significances to the results. The REFLEX and BCS samples
as well as the radio halo/relic and cooling flow samples are described
in Sect.\,\ref{SAMPLES}. Further tests and illustrations of the
morphological classifications are given in Sect.\,\ref{FTESTS}. The
observed frequency distributions of clusters with substructure are
discussed in Sect.\,\ref{SELECT}. The substructure occurrence rates
(SORs) show a comparatively strong sensitivity to the number of X-ray
photons used for substructure analyses. A simple method to reduce this
effect is described.

To proceed further, two methods seem useful. The first possibility is
the application of our substructure tests to simulated images where
selection effects introduced by observation and data reduction are
taken into account. In addition, the comparison with simulations is
expected to yield statistical information about the mass spectrum of
the merger masses to which our substructure tests are sensitive at
various redshifts and X-ray fluxes. This work is postponed to a later
paper. The second possibility is the relative comparison of subsamples
extracted from the combined REFLEX$+$BCS catalogs in a manner that
selection effects partially cancel out. The subsequent sections of
this paper are devoted to this kind of analysis.

One important result obtained with the latter strategy is the
discovery of a substructure-density relation, analogous to the
morphology-density relation for galaxies (Sect.\,\ref{MORPH}). Other
interesting results are obtained from the relative comparison of
substructure significance distributions of halo/relic, cooling flow,
and REFLEX$+$BCS clusters (Sect.\,\ref{OBS1}). Here we create
reference subsamples drawn randomly from the REFLEX$+$BCS sample in
such a way that the reference samples have basically the same
`sensitivity' for substructure detection as the halo/relic and cooling
flow samples. With this method we can compare for the first time the
substructuring and elongation behaviour of specific cluster types with
a representative large-area cluster sample. A general discussion of
the results is given in Sect.\,\ref{DISCUSS}.

All computations assume Friedmann-Lema\^{\i}tre world models with zero
pressure, the Hubble constant $H_0=50\,{\rm km}\,{\rm s}^{-1}\,{\rm
Mpc}^{-1}$, the density parameter $\Omega_0=1.0$, and the cosmological
constant $\Omega_\Lambda=0$.

\section{Statistical tests for substructure}\label{METHODS}

\subsection{General considerations}\label{GENERAL}

In X-rays, quantitative analyses of substructure are based on detailed
fits of elliptical models or wavelets to X-ray isophotes, studies of
ellipticities, center-shifts, power ratios, etc. (see, e.g., the
review in Buote 2001). However, representative SORs can only be
obtained with large cluster samples. The RASS-3 X-ray images of our
cluster sample which can be used for this purpose have on average 359
with up to 6\,829 X-ray photons (in the ROSAT hard energy band, see
below) within an aperture of 1\,Mpc radius. Therefore, more robust
tests for substructure are needed. Unfortunately, the relation between
substructure as defined by robust tests and a physical quantification
of substructure is less direct and a larger effort is needed for the
physical interpretation of the results.  The link between SORs and
theoretical merger rates can be obtained when the mass scales of the
subclumps and the time scales needed for the merged cluster to reach
dynamical equilibrium are known. Many observational effects become
apparent when statistical samples are analysed (see
Sects.\,\ref{SELECT}, \ref{TRUE}) which complicate the interpretation
of SORs. After proper correction, semi-analytic Press Schechter-like
theories as presented in, e.g., Lacey \& Cole (1993) can in principle
be applied to understand the merger rates within a cosmological
context.

However, the relation between observed SORs and true substructure
becomes secondary when SORs obtained for different subsamples drawn
from the same parent distribution are compared.  In this case the
tests define substructure more operationally as that what they
measure, and they can serve as a mere link between different cluster
types as, for example, clusters located in high and low-density
regions (Sect.\,\ref{MORPH}) or halo/relic clusters, cooling flow
clusters (Sect.\,\ref{OBS1}).

N-body simulations of merging clusters of galaxies favour three tests
for the analysis of two-dimensional point distributions (Pinkney et
al. 1996). The tests described in Sects.\,\ref{SECT_BETA} to
\ref{SECT_LEE} are sensitive to different types of substructure and
are thus ideally suited for the detection of a large variety of
different merger events. We use the three tests and translate them to
the case of two-dimensional X-ray images as extracted from the RASS-3
fields. Sect.\,\ref{FTESTS} compares substructures as defined in the
present paper under realistic conditions with the results obtained by
other research groups with different methods.

\subsection{X-ray survey data}\label{DATA}

Before the substructure statistics are described the basic properties
of the X-ray material used for the analysis are summarized. This will
explain some of the constraints already imposed by the observational
data. The 1\,378 RASS-3 fields used for substructure analyses cover
the whole sky with an {\it averaged} spatial resolution as given by
the half power radius of 96 arcsec at 1\,keV. Substructure on scales
larger than approximately $2$\,Mpc\,$\cdot z$ (e.g., 200\,kpc at
$z=0.1$) is thus resolved. Within this precision, the point spread
function is constant over each RASS field and can be regarded as
circular symmetric (G. Boese, private communication). Each field has a
size of $6.4 \times 6.4\,{\rm deg}^2$ and overlaps at least 0.23\,deg
with adjacent fields. The main advantage of RASS-3 compared to RASS-2
is that its less stringent constraints on the attitude solutions yield
a larger number of accepted X-ray photon events (on average about 5
percent, and is essential only in certain parts of the sky) resulting
in a higher signal-to-noise without a significant increase of the
measurement errors of the individual photons.  The advantages of
RASS-2 against RASS-1 are discussed in Voges et al. (1999).

Each of the $N_{\rm ph}$ photons detected in the direction of a
cluster of galaxies is characterized by the following quantities.  The
sky pixel coordinates, $(x_i,y_i)$, given for the $i$th photon in
units of 0.5 arcsec, and the photon energy, $E_i$, given in the form
of a Pulse Height Amplitude (PHA) channel number, are contained in the
photon event file of each RASS-3 field. The exposure time, $t_i$, used
to weight each photon, is determined with the exposure map, giving for
each RASS-3 field the survey exposure times, corrected for vignetting
and the effect of the shadowing of the support structure of the PSPC
window with a spatial binning of 45 arcsec.  We do not differentiate
between source and background photons because no distinction can be
made on the photon-by-photon basis.

Only X-ray photons selected from the RO\-SAT hard energy band (PHA
channels 52 to 201 corresponding to about 0.5-2.0\,keV) are used for
all reductions. This reduces the soft X-ray background by a factor of
4, but still keeps 60 to 100 percent (depending on the interstellar
column density) of the cluster emission. Moreover, the majority of
soft RASS-3 sources superposed onto the cluster images are suppressed,
so that the signal-to-noise for cluster detection is highest.

Since comparatively large samples of clusters have to be analyzed the
computer-intensive substructure tests had to be simplified (see
below). The tests are performed within one aperture with the metric
radius of $1.0\,{\rm Mpc}$ centered on the X-ray intensity-weighted
center of the cluster. The iterative determination of the cluster
center terminates when a formal accuracy of 2 arcsec is reached (for
more details see B\"ohringer et al. 2000, 2001).

\subsection{The $\beta$ test}\label{SECT_BETA}
 
The $\beta$ statistic compares the surface number density, $\mu_i$,
measured in the $i$th radial segment with the density, $\mu_{oi}$,
measured in the diametrically opposite segment. The number densities
are obtained by weighting each photon with the corresponding exposure
time. The final $\beta$ value is the average of the ratios obtained
over all independent $N_{\rm s}$ radial segments (typically 8):
\begin{equation}\label{BETA}
\beta\,=\,\frac{1}{N_{\rm s}}\,\sum_{i=1}^{N_{\rm s}}
\log_{10}\left(\frac{\mu_i}{\mu_{oi}}\right)\,.
\end{equation}
This test is a simplified version of a method developed by West,
Oemler, \& Dekel (1988).

As noted in Pinkney et al. (1996) the $\beta$ value (in its original
definition) is sensitive to deviations from mirror symmetry,
independent of the actual elongation of the target. The statistic
becomes rather ineffective as the mass ratio of the sub-components
approaches $1:1$. We expect similar properties of the presently
implemented simplified version. Note that the application of radial
segments instead of individual photon coordinates is used to make the
test more robust and to save computing time although this might
desensitize the test. The actual sensitivity of the $\beta$ parameter
is tested with RASS-3 images in Sect.\,\ref{FTESTS}.

\subsection{The Lee test}\label{SECT_LEE}
 
In the Lee statistic the photons are projected onto lines with given
inclination angles, $\phi$ (12 directions in the present
implementation). In general, for all $\phi$ and for all partitions of
the set of photons, the `within-class' scatter, $\sigma_{\rm
L}+\sigma_{\rm R}$, and the `between-class' scatter, $\sigma_{\rm T}$,
are determined and their ratios, $\sigma_{\rm T}/(\sigma_{\rm
L}+\sigma_{\rm R})$, maximized. We use the test based on the technique
of maximum likelihood in the form
\begin{equation}\label{LEE1}
L\,=\,\frac{L_{\rm max}}{L_{\rm min}}\,=\,\frac{
 {\rm max}_\phi\left\{L(\phi)\right\}}
{{\rm min}_\phi\left\{L(\phi)\right\}}\,,
\end{equation}
where 
\begin{equation}\label{LEE2}
L(\phi)\,=\,{\rm max}_{\rm \{partitions\}}\left(\frac{\sigma_{\rm
T}}{\sigma_{\rm L}+\sigma_{\rm R}}-1\right)
\end{equation}
gives the maximum likelihood obtained for the $(N_{\rm ph}-1)$
partitions. The partitions are obtained by dividing the total set of
photon coordinates (${\rm T}$) projected onto a line with a given
$\phi$ into a right (${\rm R}$) and a left (${\rm L}$) subgroup with
the corresponding scatters
\begin{equation}\label{LEE3}
\sigma_\alpha\,=\sum_{j=1}^{K_\alpha} w_j\,(x_j-\bar{x}_\alpha)^2\,,
\quad\quad \bar{x}_\alpha\,=\,\frac{\sum_{j=1}^{K_\alpha}
w_j\,x_j}{\sum_{j=1}^{K_\alpha} w_j}\,,
\end{equation}
where $\alpha={\rm T},{\rm R},{\rm L}$, the weighting factor for the
$i$th photon is $w_j=1/t_j$, and $K_\alpha$ is the actual number of
projected points in the ${\rm T},{\rm R},{\rm L}$ samples.  More
details can be found in Fitchett et al. (1988). Fitchett \& Webster
(1987) successfully applied the method for substructure detection in
the core of the Coma galaxy cluster.

The likelihood value, $L$, is most sensitive if two substructure
components are present, especially when they are very compact. It is
not sensitive to any elongations and loses sensitivity if more than
two subclumps are visible. Compared to the other two tests the LEE
statistic is thus the most conservative.

\subsection{The Fourier elongation test}\label{SECT_FEL}
 
Deviations from circular symmetry might indicate merger events (e.g.,
Roettiger et al. 1997, see also Buote 2001), although relaxed clusters
can show significant ellipticities too as illustrated in
Sect.\,\ref{FTESTS}.  However, recent simulations of Thomas et
al. (1998) and others (see Sect.\,\ref{INTRO}) clearly show a
dependency of the frequency distribution of cluster major axial ratios
on cosmology, where the rounder X-ray isophotes in the low-$\Omega_0$
models might result primarily from the scarcity of recent mergers
(Evrard et al. 1993). We thus regard elongation as a useful
cosmological quantity although its relation to substructure appears to
be more complicated.

Following Rhee, van Haarlem \& Katgert (1991) the azimuthal number
counts are approximated to first order by a constant density,
modulated by a double sine. Under this assumption the normalized
amplitude of this modulation,
\begin{equation}\label{FEL1}
{\rm FEL}\,=\,\sqrt{\frac{2\left(S^2+C^2\right)}{N_0}}\,,
\end{equation}
gives a measure of the elongation strength. The parameters $S$ and $C$
are defined by equation (7) in Pinkney et al. (1996). The position
angle is then ${\rm PA}=0.5\arctan(S/C)$. We translate (\ref{FEL1}) to the
unbinned case, where each photon is weighted by its exposure time,
\begin{equation}\label{FEL2}
S\,=\,\sum_{i=1}^{N_{\rm
ph}}\,\frac{2x_iy_i}{t_i\,(x_i^2+y_i^2)}\,,\quad\quad
C\,=\,\sum_{i=1}^{N_{\rm
ph}}\,\frac{x_i^2-y_i^2}{t_i\,(x_i^2+y_i^2)}\,,
\end{equation}
with $N_0=\sum_i\,1/t_i$. The N-body simulations show that FEL is the
most sensitive of the three tests used to check for substructure if
elongation is to be considered as substructure.

\subsection{Statistical significances}\label{SIGNIF}

In the following the method is described which is used to compute the
probabilities (statistical significances $S$) that the actual values
of the substructure parameters described in Sects.\,\ref{SECT_BETA} to
\ref{SECT_LEE} could be obtained just by chance from an X-ray image
satisfying the null hypothesis of a circular symmetric,
mirror-symmetric, and unimodal surface brightness distribution. The
statistical significance, $S$, thus corresponds to the confidence
probability, $(1-S)$, that the null hypothesis can be rejected.

The significance values, $S_\beta$, $S_{\rm FEL}$, $S_{\rm LEE}$, are
computed by comparing the $\beta$, ${\rm FEL}$, and $L$ values
obtained for the programme cluster with the corresponding values
obtained with a large set of unstructured photon distributions
(replicants) derived from the same cluster. For the
position-independent, circular symmetric average point spread function
of RASS-3 (see Sect.\,\ref{DATA}), these smooth and symmetric
distributions can be obtained by azimuthal randomization (West et
al. 1988), keeping the radial distances of the original X-ray photons,
but assigning random angular coordinates without the need to choose a
specific (model) cluster profile.

The significances are computed as follows. In the first step $\beta$,
${\rm FEL}$, and $L$ are computed for the programme cluster. In the
next step the photon distribution is azimuthally randomized, and for
each randomized cluster, $\beta'$, ${\rm FEL}'$, and $L'$ are computed
again. The primes denote values of the statistics obtained with the
randomized distribution. As seen above the statistics are normalized
in such a way that larger values correspond to larger
substructures. Therefore, the number of times the randomized
distribution gives a value larger than that obtained for the programme
cluster provides an estimate of the probability, $S$, that the actual
substructure value can be obtained just by chance from a cluster
fulfilling the null hypothesis. Notice that small values of $S$
correspond to clusters with substructure and elongation.

To be more specific, the substructure analyses are performed and
statistical significances are computed for one aperture with the
metric radius of $1.0\,{\rm Mpc}$. This metric scale is transformed
into an angular scale with the cluster redshift using the angular
diameter distance. The choice of this radius guarantees that in the
majority of cases the outer significance radii of the X-ray images,
defined below, include the aperture used to evaluate substructure. For
larger apertures, contributions from neighbouring clusters and chance
superposition of background sources become more important. The
significance radius is defined in B\"ohringer et al. (2000, 2001) as
the point where the increase in the $1\sigma$ flux error is larger
than the increase of the cumulative source count rate after background
subtraction. For the $\beta$ and ${\rm FEL}$ statistic, 2\,000
replicants are used for each cluster, for the most time-consuming LEE
statistic 400 replicants.

\begin{figure*}
\vspace{0.0cm} \centerline{\hspace{0.0cm}
\psfig{figure=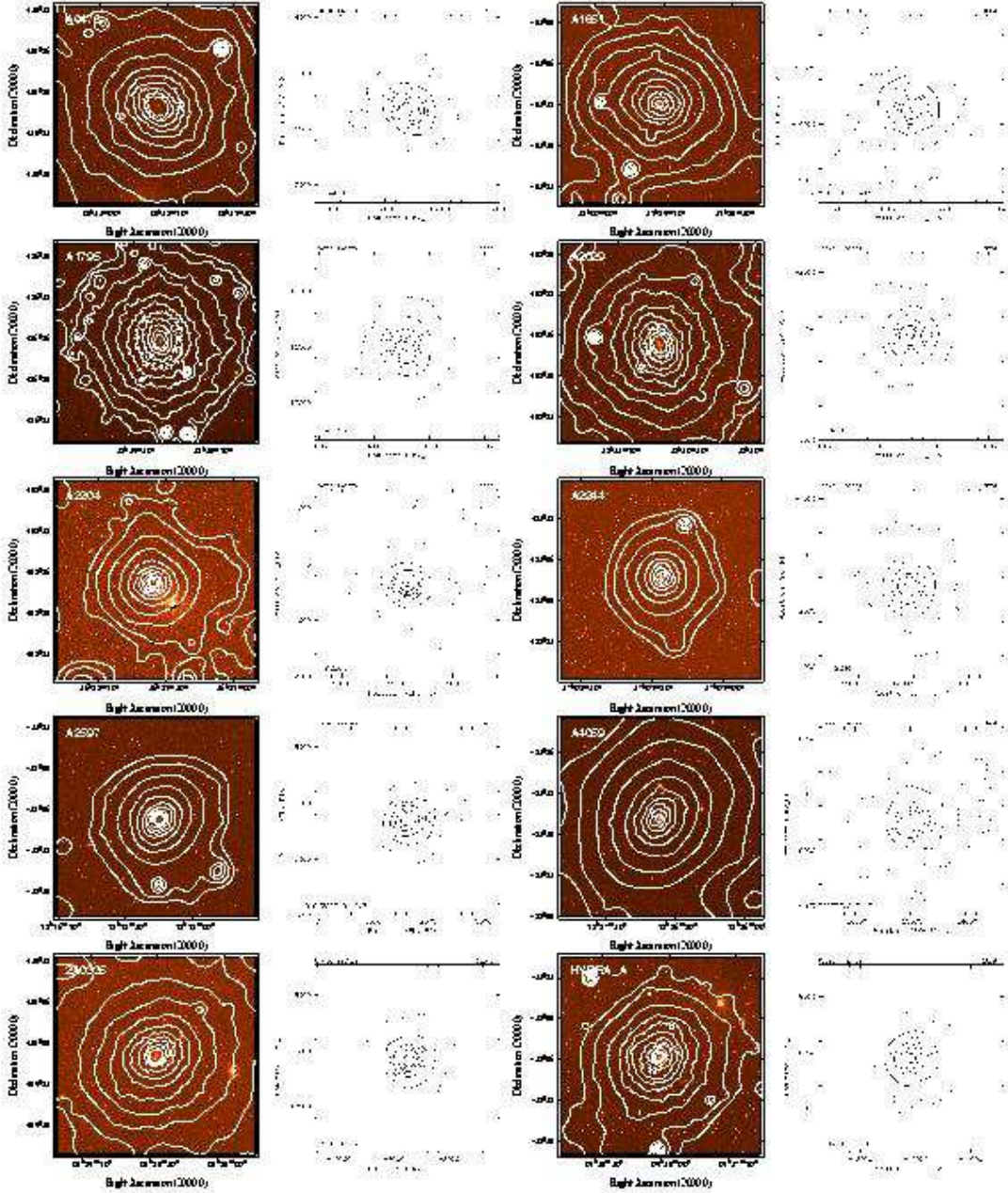,height=20.0cm}
} \vspace{-2.00cm}
\caption{\small Comparison of iso-surface brightness contours of deep
ROSAT PSPC pointings (first and third columns) and corresponding
RASS-3 images (second and fourth columns) of regular clusters. The
pointings are superposed on Digital Sky Survey optical images. Source
characterizations are given in Tab.\,\ref{TAB_CHAR1} (clusters 1-10).}
\label{FIG_ARRAY1}
\end{figure*}

\begin{figure*}
\vspace{0.0cm} \centerline{\hspace{0.0cm}
\psfig{figure=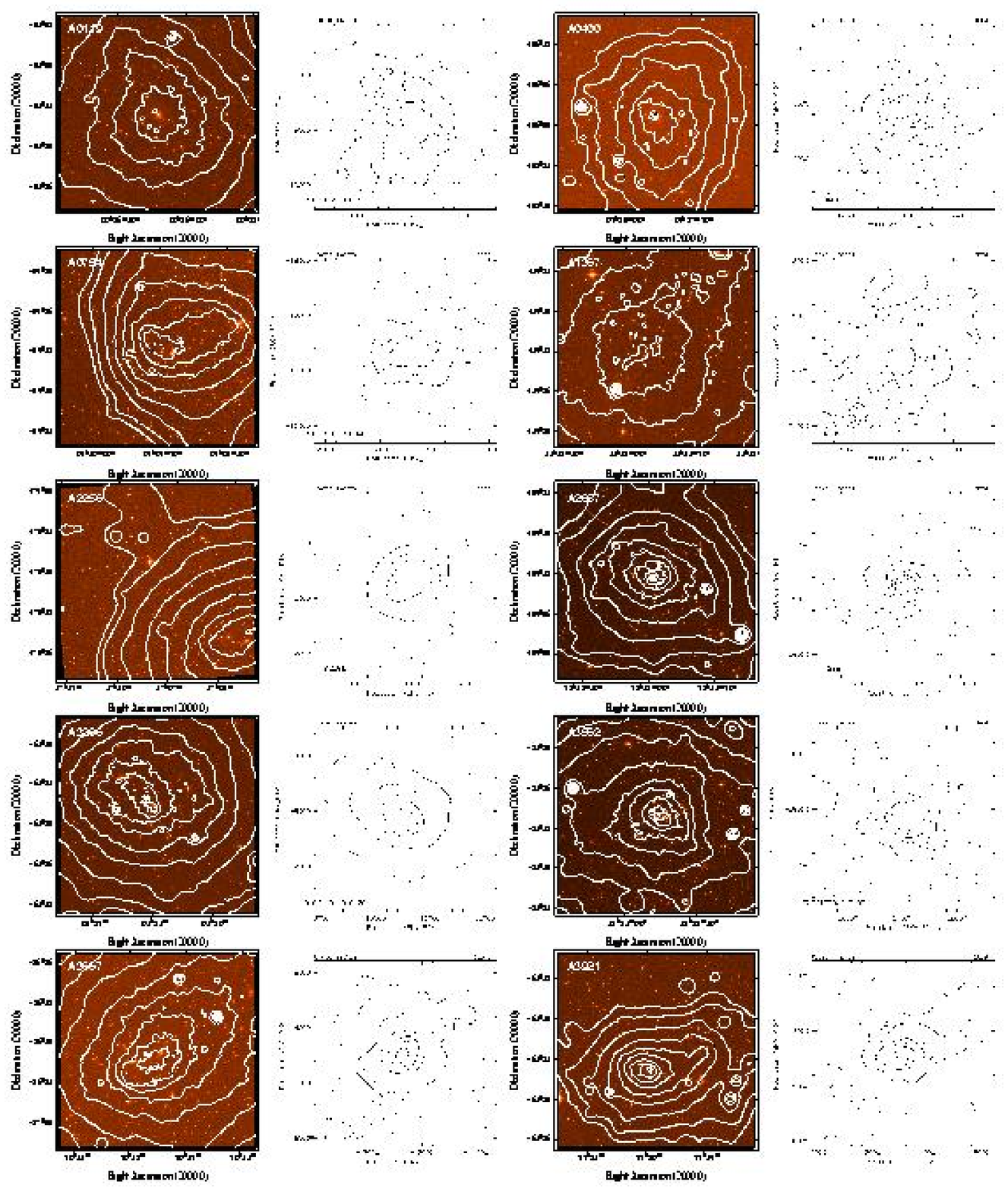,height=20.0cm}
} \vspace{-2.00cm}
\caption{\small Comparison of iso-surface brightness contours of deep
ROSAT PSPC pointings (first and third columns) and corresponding
RASS-3 images (second and fourth columns) of substructured
clusters. The pointings are superposed on Digital Sky Survey optical
images. Source characterizations are given in Tab.\,\ref{TAB_CHAR2}
(clusters 11-20).} \label{FIG_ARRAY2}
\end{figure*}

\section{Cluster samples}\label{SAMPLES}

The REFLEX sample (B\"ohringer et al. 2001) includes 452 clusters (449
with redshifts), detected in the RASS-2 database (Voges et al.  1999)
south of the Declination $+2.5$\,deg and down to the nominal X-ray
flux limit of $3.0\times 10^{-12}\,{\rm erg}\,{\rm s}^{-1}\,{\rm
cm}^{-2}$ in the RO\-SAT energy band ($0.1$-$2.4$\,keV). REFLEX covers
4.24\,sr excluding the area $\pm 20$\,deg around the galactic plane
and 0.0987\,sr around the Magellanic Clouds and is basically
restricted to redshifts $z\le 0.3$ (a few clusters reach
$z=0.45$). The Growth Curve Analysis method of B\"ohringer et al.
(2000, 2001) is used to compute source positions, fluxes, angular
extents, etc. (source characterization). The catalog gives unabsorbed
X-ray fluxes with statistical errors between 10 and 20 percent.
Several tests indicate a high overall completeness of the sample (at
least 90 percent) with an upper limit of 10 percent of the clusters
with fluxes significantly contaminated by active galactic nuclei
(AGN).

The BCS sample (Ebeling et al. 1998) includes 201 clusters
(statistical sample), detected north of the Declination
$\delta=0$\,deg, excluding the area $\pm 20$\,deg around the galactic
plane using the RASS-1 data. The formal flux limit of the resulting
sample is $4.4\times 10^{-12}\,{\rm erg}\,{\rm s}^{-1}\,{\rm cm}^{-2}$
($0.1$-$2.4$\,keV).  The Voronoi Tesselation and Percolation method
(first version of the method described in Ebeling \& Wiedenmann 1993)
is used to obtain unabsorbed X-ray fluxes. The authors estimated a
sample completeness of 90 percent for the 201 BCS clusters with $z\le
0.3$. Some artificial fluctuations in cluster number density might
occur because the clusters are not sampled homogeneously over the
total survey area given in Ebeling et al. (1998). For the present
investigation the effects can be neglected because only
large-amplitude fluctuations ($>10$ precent) are discussed.

There are basically two motivations for using a combined sample of
REFLEX and BCS clusters for our substructure analysis. (1) It is
expected that the results of substructure analyses are improved when
more X-ray photons are available (see Sect.\,\ref{SELECT}). Combining
the REFLEX and BCS catalogues to an all-sky sample and cutting it at
the flux limit of the brighter BCS sample yields the largest number of
X-ray images with high numbers of X-ray photons. (2) Due to the fact
that at low redshifts the northern BCS sample might be dominated by
the local supercluster and its extension, and the southern REFLEX
sample by the southern void (e.g., Schuecker et al. 2001), the merged
REFLEX and BCS catalogue is expected to provide a more representative
sample.

To ensure an homogeneous handling of the data we exclude the Virgo
cluster because of its large angular extent. Furthermore, A689, Z3179,
and RXCJ1212.3-1817 are excluded because of their low ROSAT exposure
times (too few photons), A1678 and RXCJ0532.9-3701 which are located
too close to the RASS-3 field edge, and three further REFLEX clusters
where redshift information is not yet available. The combined sample
referred to as RE\-FLEX$+$BCS contains 470 clusters (excluding the BCS
clusters in common with REFLEX) with a formal flux limit of $4.4\times
10^{-12}\,{\rm erg}\,{\rm s}^{-1}\,{\rm cm}^{-2}$. Although BCS and
REFLEX fluxes are computed with different algorithms on different RASS
versions no problems are expected because each version has its own
calibrations. 

A sample of 53 halo and relic clusters (including six uncertain cases)
compiled by Feretti et al. (in preparation) is used to study the
statistical properties of their X-ray morphology. In the following we
use the combined halo/relic sample as mentioned in
Sect.\,\ref{INTRO}. Deep pointings obtained mainly with RO\-SAT
suggest that eventually all of these clusters show distorted X-ray
surface brightness distributions. Presently no information about the
completeness of this sample is available. In the present investigation
RASS-3 data are used for their analysis so that the substructure tests
can be directly compared to the statistically representative
REFLEX$+$BCS reference sample.

A sample of 22 clusters with large cooling flow signatures is selected
from the list presented in Peres et al. (1998). Their total sample of
55 clusters with fluxes above $1.7\times 10^{-11}\,{\rm erg}\,{\rm
s}^{-1}\,{\rm cm}^{-2}$ in the $2$-$10$\,keV energy band is selected
from observations with the {\it Einstein} and EXOSAT observatories,
and the HEAO-1 and {\rm Ariel}\,V satellites by Edge et al. (1990) and
is found to be `satisfactorily complete'. The selected cooling flow
clusters have mass deposition rates larger than $100\,M_\odot\,{\rm
yr}^{-1}$ as determined by a surface brightness deprojection technique
using RO\-SAT\,PSPC and HRI pointed observations. Notice that for the
present investigation only the cooling flow signature of a steep
increase of the central X-ray surface brightness matters, and that a
possibly new interpretation of cooling flows due to the non-detection
of cooling gas with temperatures below about 3\,keV in XMM spectra
(Peterson et al. 2001) is secondary. The majority of the clusters in
the sample have $z\le 0.1$.

\section{Verification of the method}\label{FTESTS}

\begin{table*}
\caption{Characterization of the regular galaxy clusters shown in
Fig.\,\ref{FIG_ARRAY1}. The mass deposition rates are in units of
solar mass per year as obtained from ROSAT PSPC/HRI pointings, $S_{\rm
LEE}$, $S_\beta$, $S_{\rm FEL}$ are the statistical significances
obtained with the RASS-3 images within circles of radius $R$ in arcsec
corresponding to a metric radius of $1.0\,{\rm Mpc}$, position angles,
${\rm PA}$ in degrees as obtained with FEL and surface
brightness-weighted moments.}
\begin{center}
\begin{tabular}{rlll}
Cluster  & Characterization & Reference \\ 
\hline
1 & A478     & smooth             & Buote \& Tsai (1996) \\
  &          & no centroid shift  & Mohr et al. (1995) \\
  &          & $\dot{M}=520-616$  & Peres et al. (1998) \\
  &          & $R=8'$: $S_{\rm LEE}=0.28$, $S_\beta=0.76$, $S_{\rm FEL}=0.00$, ${\rm PA}=58/57$ \\
2 & A1651    & mostly smooth$+$ regular & Buote \& Tsai (1996) \\
  &          & $\dot{M}=138$      & Peres et al. (1998) \\
  &          & $R=8'$: $S_{\rm LEE}=0.56$, $S_\beta=0.47$, $S_{\rm FEL}=0.53$, ${\rm PA}=100/105$ \\
3 & A1795    & smooth $+$ regular & Buote \& Tsai (1996) \\
  &          & $\dot{M}=381/488$  & Peres et al. (1998) \\
  &          & $R=10'$: $S_{\rm LEE}=0.15$, $S_\beta=0.64$, $S_{\rm FEL}=0.03$, ${\rm PA}=20/12$ \\
4 & A2029    & regular $+$ smooth   & Buote \& Tsai (1996) \\
  &          & $\dot{M}=554/556$  & Peres et al. (1998) \\
  &          & $R=9'$: $S_{\rm LEE}=0.87$, $S_\beta=0.07$, $S_{\rm FEL}=0.81$, ${\rm PA}=169/176$ \\
5 & A2204    & smooth             & Buote \& Tsai (1996) \\
  &          & $\dot{M}=843/852$  & Peres et al. (1998) \\
  &          & $R=5'$: $S_{\rm LEE}=0.59$, $S_\beta=0.45$, $S_{\rm FEL}=0.17$, ${\rm PA}=77/92$ \\
6 & A2244    & regular            & Buote \& Tsai (1996) \\
  &          & $\dot{M}=244$      & Peres et al. (1998) \\
  &          & $R=7'$: $S_{\rm LEE}=0.13$, $S_\beta=0.10$, $S_{\rm FEL}=0.02$, ${\rm PA}=6/174$ \\
7 & A2597    & very symm. $+$ smooth& Buote \& Tsai (1996) \\
  &          & $\dot{M}=276/271$  & Peres et al. (1998) \\
  &          & $R=8'$: $S_{\rm LEE}=0.16$, $S_\beta=0.49$, $S_{\rm FEL}=0.02$, ${\rm PA}=93/112$ \\
8 & A4059    & smooth, single-comp. & Buote \& Tsai (1996) \\
  &          & $\dot{M}=130$      & Peres et al. (1998) \\
  &          & $R=13'$: $S_{\rm LEE}=0.96$, $S_\beta=0.30$, $S_{\rm FEL}=0.01$, ${\rm PA}=127/130$ \\
9 & 2A0335   & $\dot{M}=242/325$  & Peres et al. (1998) \\
  &          & $R=18'$: $S_{\rm LEE}=0.65$, $S_\beta=0.06$, $S_{\rm FEL}=0.52$, ${\rm PA}=132/146$ \\
10& Hydra A  & regular            & Buote \& Tsai (1996) \\
  &          & $\dot{M}=298/264$ & Peres et al. (1998) \\
  &          & $R=12'$: $S_{\rm LEE}=0.35$, $S_\beta=0.37$, $S_{\rm FEL}=0.66$, ${\rm PA}=140/126$ \\
\hline
\end{tabular}
\end{center}
\label{TAB_CHAR1}
\end{table*}

\begin{table*}
\caption{Characterization of the substructured galaxy clusters shown
in Fig.\,\ref{FIG_ARRAY2}. The mass deposition rates are in units of
solar mass per year as obtained from ROSAT PSPC/HRI pointings, $S_{\rm
LEE}$, $S_\beta$, $S_{\rm FEL}$ are the statistical significances
obtained with the RASS-3 images within circles of radius $R$ in arcsec
corresponding to a metric radius of $1.0\,{\rm Mpc}$, position angles
${\rm PA}$ in degrees as obtained with FEL and surface
brightness-weighted moments.}
\begin{center}
\begin{tabular}{rlll}
Cluster  & Characterization & Reference \\ 
\hline
11& A119     & emission tail      & Buote \& Tsai (1996) \\
  &          & opt. substructured & Kriessler \& Beers (1997) \\
  &          & centroid shift     & Mohr et al. (1995) \\
  &          & $\dot{M}=0$        & Peres et al. (1998) \\
  &          & $R=14'$: $S_{\rm LEE}=0.84$, $S_\beta=0.00$, $S_{\rm FEL}=0.13$, ${\rm PA}=17/27$ \\
12& A400     & irregular          & Buote \& Tsai (1996) \\
  &          & centroid shift     & Mohr et al. (1995) \\
  &          & $R=25'$: $S_{\rm LEE}=0.02$, $S_\beta=0.34$, $S_{\rm FEL}=0.11$, ${\rm PA}=176/155$ \\
13& A754     & clearly off center & Buote \& Tsai (1996) \\
  &          & opt. substructured & Kriessler \& Beers (1997) \\
  &          & centroid shift     & Mohr et al. (1995) \\
  &          & $\dot{M}=0/2$      & Peres et al. (1998) \\
  &          & Radio halo $+$ relic& Cohen et al. (2001) \\
  &          & Cold Front         & Markevitch et al. 2001 \\
  &          & $R=12'$: $S_{\rm LEE}=0.00$, $S_\beta=0.00$, $S_{\rm FEL}=0.00$, ${\rm PA}=101/101$ \\
14& A1367    & centroid shift     & Mohr et al. (1995) \\
  &          & $\dot{M}=8/0$      & Peres et al. (1998) \\
  &          & $R=28'$: $S_{\rm LEE}=0.00$, $S_\beta=0.00$, $S_{\rm FEL}=0.00$, ${\rm PA}=146/145$ \\
15& A2256    & unrelaxed          & Buote \& Tsai (1996) \\
  &          & elliptical         & Jones \& Forman (1999) \\
  &          & substructured core & Mohr et al. (1995) \\
  &          & $\dot{M}=0/0$      & Peres et al. (1998) \\
  &          & Radio halo $+$ relic& Feretti (1999) \\
  &          & $R=11'$: $S_{\rm LEE}=0.00$, $S_\beta=0.03$, $S_{\rm FEL}=0.00$, ${\rm PA}=128/128$ \\
16& A2657    & bi-modal           & Buote \& Tsai (1996) \\
  &          & opt. substructured & Kriessler \& Beers (1997) \\
  &          & $R=15'$: $S_{\rm LEE}=0.02$, $S_\beta=0.23$, $S_{\rm FEL}=0.00$, ${\rm PA}=77/85$ \\
17& A3266    & substructured core & Mohr et al. (1995) \\
  &          & strong substructure& Kolokotronis et al. (2000) \\
  &          & $\dot{M}=0/3$      & Peres et al. (1998) \\
  &          & $R=11'$: $S_{\rm LEE}=0.00$, $S_\beta=0.00$, $S_{\rm FEL}=0.00$, ${\rm PA}=65/63$ \\
18& A3562    & distorted/not ellip.& Buote \& Tsai (1996) \\
  &          & $\dot{M}=37$       & Peres et al. (1998) \\
  &          & $R=12'$: $S_{\rm LEE}=0.68$, $S_\beta=0.02$, $S_{\rm FEL}=0.60$, ${\rm PA}=94/61$ \\
19& A3667    & elongated $+$ distort.& Buote \& Tsai (1996) \\
  &          & apparent bi-modal  & Kolokotronis et al. (2000) \\
  &          & $\dot{M}=0$        & Peres et al. (1998) \\
  &          & Cold Front         & Vikhlinin et al. (2001) \\
  &          & 2 radio relics     & Feretti (1999) \\
  &          & $R=11'$: $S_{\rm LEE}=0.00$, $S_\beta=0.80$, $S_{\rm FEL}=0.00$, ${\rm PA}=133/113$ \\
20& A3921    & irregular/merger   & Buote \& Tsai (1996) \\
  &          & apparent bi-modal  & Kolokotronis et al. (2000) \\
  &          & $R=7'$: $S_{\rm LEE}=0.02$, $S_\beta=0.88$, $S_{\rm
FEL}=0.00$, ${\rm PA}=99/102$ \\
\hline
\end{tabular}
\end{center}
\label{TAB_CHAR2}
\end{table*}

\begin{table*}
\caption{Selected \gcs\ clusters and their statistical significances
for the LEE, $\beta$, and FEL statistics for regular (Reg), possibly
substructured (Poss), and substructured (Sub) clusters. The
significances are obtained from RASS-3 images, the classification of
the clusters is taken from the literature. Mean significances (Mean1)
and their formal standard deviation (Std1) as well as the mean
significances (Mean2) and their standard deviation (Std2) obtained for
the minimum values of ($S_{\rm LEE},S_\beta$) are given in the last
rows (eq.\,\ref{MIN}). Notice that A85 has a substructure just outside
the 1\,Mpc aperature and is thus regarded as possibly substructured.}
\begin{center}
\begin{tabular}{|lccc|lccc|lccc|}
 & Reg & & & & Poss & & & & Sub & &  \\
Name&$S_{\rm LEE}$&$S_\beta$&$S_{\rm FEL}$& 
Name&$S_{\rm LEE}$&$S_\beta$&$S_{\rm FEL}$& 
Name&$S_{\rm LEE}$&$S_\beta$&$S_{\rm FEL}$\\
\hline
A478   &0.28&0.76&0.00&A85  &0.48&0.21&0.00&A119   &0.84&0.00&0.13\\
A1651  &0.56&0.47&0.53&A133 &0.59&0.11&0.13&A400   &0.02&0.34&0.11\\
A1795  &0.15&0.64&0.03&A401 &0.07&0.22&0.03&A496   &0.01&0.14&0.00\\
A2029  &0.87&0.07&0.81&A2063&0.70&0.01&0.77&A754   &0.00&0.00&0.00\\
A2052  &0.80&0.02&0.06&A3158&0.01&0.13&0.00&A1367  &0.02&0.00&0.00\\
A2204  &0.59&0.45&0.17&A3558&0.07&0.06&0.00&A2142  &0.04&0.00&0.05\\
A2244  &0.13&0.10&0.02&     &    &    &    &A2147  &0.00&0.00&0.00\\
A2589  &0.01&0.39&0.00&     &    &    &    &A2255  &0.00&0.00&0.00\\
A2597  &0.16&0.49&0.02&     &    &    &    &A2256  &0.00&0.03&0.00\\
A3112  &0.39&0.90&0.02&     &    &    &    &A2634  &0.02&0.01&0.02\\
A3581  &0.13&0.96&0.13&     &    &    &    &A2657  &0.02&0.23&0.00\\
A4038  &0.00&0.95&0.00&     &    &    &    &A3266  &0.00&0.00&0.00\\
A4059  &0.96&0.30&0.01&     &    &    &    &A3376  &0.00&0.00&0.00\\
MKW3s  &0.24&0.44&0.21&     &    &    &    &A3391  &0.00&0.46&0.00\\
Hydra-A&0.35&0.37&0.66&     &    &    &    &A3395  &0.00&0.00&0.00\\
S1101  &0.73&0.11&0.62&     &    &    &    &A3562  &0.68&0.02&0.60\\
2A0335 &0.65&0.06&0.52&     &    &    &    &A3571  &0.00&0.92&0.00\\
       &    &    &    &     &    &    &    &A3667  &0.00&0.80&0.00\\
       &    &    &    &     &    &    &    &A3921  &0.02&0.88&0.00\\
       &    &    &    &     &    &    &    &RXJ2344&0.04&0.02&0.06\\
\hline
Mean1  &0.412&0.440&0.224& &0.320&0.123&0.155& &0.086&0.193&0.049\\
Std1   &0.075&0.077&0.068& &0.124&0.034&0.125& &0.052&0.071&0.030\\
Mean2  &0.194&    &    &     &0.078&    &    &       &0.006&    &    \\
Std2   &0.038&    &    &     &0.031&    &    &       &0.002&    &    \\
\hline
\end{tabular}
\end{center}
\label{TAB_CHAR3}
\end{table*}

The methods used for substructure detection were analyzed and tested
in detail with numerical N-body simulations (Pinkney et
al. 1996). Here we want to illustrate the types of substructures to
which the three substructure tests are sensitive under realistic
conditions comparing deep ROSAT pointed observations and the
corresponding RASS-3 images of clusters with known morphology and
substructure behaviour (see Sect.\,\ref{TEST1}). In addition, we want
to study the relation between the substructure tests and the Jones \&
Forman classification scheme to allow a reference to a similar
systematic survey project (see Sect.\,\ref{TEST2}). It will be seen
that the three substructure statistics provide a quantitative
morphological classification of RASS-3 images which has a close
relation to substructure as defined and quantified with other methods
and data.

\subsection{Comparison with deep ROSAT PSPC pointings}\label{TEST1}

Figures \ref{FIG_ARRAY1} and \ref{FIG_ARRAY2} show a collection of 10
regular and 10 substructured `prototype' clusters as obtained with
deep ROSAT PSPC pointings (1st and 3rd columns) and the corresponding
RASS-3 images (2nd and 4th columns) used for our substructure
analyses. Literature information and the significances obtained with
our substructure tests using RASS-3 data within a metric aperture of
$1.0\,{\rm Mpc}$ are summarized in Tabs.\,\ref{TAB_CHAR1} and
\ref{TAB_CHAR2}. From the literature we summarize the classifications
obtained with the power ratio technique from ROSAT PSPC pointed
observation by Buote \& Tsai (1996), cooling flow mass deposition
rates and cooling times as derived by Peres et al. (1998) using ROSAT
PSPC and HRI data, centroid variances deduced by Mohr et al. (1995),
and visual classifications of Jones \& Forman (1999), both obtained
with {\it Einstein} IPC data, ellipticities and center-of-mass shifts
as deduced from Abell$+$APM galaxy and ROSAT X-ray data by
Kolokotronis et al. (2000), and results obtained with the
adaptive-kernel technique by Kriessler \& Beers (1997) using optical
galaxies in Dressler's clusters. We also add the presence of cold
fronts as given in Markevitch et al. (2000) and Vikhlinin et
al. (2001).

The clusters shown in Fig.\,\ref{FIG_ARRAY1} are nearby $z<0.1$
clusters, all with strong cooling flow signatures and classified as
single, regular, smooth, and/or no centroid shift by the authors given
in Tabs.\,\ref{TAB_CHAR1} and \ref{TAB_CHAR2}. The $S_{\rm LEE}$
significances as obtained from the RASS-3 images range between 0.13
and 0.96, the $S_\beta$ values between 0.06 and 0.76, and the $S_{\rm
FEL}$ values between 0.0 and 0.81. Assuming a 99 percent confidence
level none of the LEE and $\beta$ tests indicate significant
substructure for these clusters. According to our substructure
statistics, clusters like A478 and A4059 show a significant
ellipticity, including A1795, A2244, A2597 on the 97 percent
confidence level (in total 5 out of the 10 clusters). If we assume
that the central $1.0\,{\rm Mpc}$ of these clusters, where the tests
are performed, are dynamically relaxed and unstructured then
ellipticity and thus FEL would turn out to be a less useful indicator
for substructure whereas $\beta$ and LEE would do.

The non-regular clusters shown in Fig.\,\ref{FIG_ARRAY2} have
negligible cooling flows, some have cold fronts. The clusters are
classified as substructured by different authors using various methods
and data (X-ray, optical). For A754, A1367, and A3266 the $\beta$,
LEE, and FEL tests all give zero probability for the null hypothesis,
thus strongly suggesting substructured RASS-3 surface brightness
distributions. For A119, A2256, and A3667 either the $\beta$ and/or
the LEE test suggest the presence of substructure. On the 98 percent
confidence level this is also true for A400, A2657, A3562, and
A3921. Therefore, basically all 10 prototype clusters shown here are
recovered as substructured on the 98 percent confidence level with the
$\beta$ and/or LEE test on RASS-3.

We found a good correspondence between the position angles, ${\rm
PA}$, as seen in the deep pointings and as obtained from the RASS-3
images (given in Tabs.\,\ref{TAB_CHAR1} and \ref{TAB_CHAR2}). The
position angles are obtained with both the FEL method and with surface
brightness-weighted moments. In basically all cases with elongation
significances $S_{\rm FEL}\le 0.01$, position angles obtained with the
FEL method and those determined with surface brightness-weighted
moments are found to be very similar within a few degrees.

Moreover, 7 of the 10 clusters (A754, A1367, A2256, A2657, A3266,
A3667, A3921) have a significant ellipticity. This suggests a
correlation between substructure and the detection of
ellipticity. However, as the analysis of the cooling flow clusters
already showed, ellipticity as a sole criterion for substructure might
not be sufficient for substructure detection. Nevertheless a final
decision whether ellipticity is regarded as substructure is not
necessary for the present investigation.

In addition to the 20 prototype clusters discussed above,
Tab.\,\ref{TAB_CHAR3} summarizes the significances obtained for 43
clusters included in the HIghest X-ray FLUx Galaxy Cluster Sample
(\gcs, see the sample description in Reiprich \& B\"ohringer 2001)
which have ROSAT PSPC pointed observations and reliable literature
information of the kind given in Tabs.\,\ref{TAB_CHAR1} and
\ref{TAB_CHAR2} about the presence or absence of substructure (see the
references given at the beginning of this section). The mean
significances (Mean1) and their formal $1\sigma$ deviations (Std1)
given in Tab.\,\ref{TAB_CHAR3} show a clear trend towards smaller
significances for the clusters expected to be substructured. This
trend is even stronger when mean and standard deviation are computed
for the minimum values
\begin{equation}\label{MIN}
S\,=\,{\rm min}\,\left\{S_{\rm LEE},S_\beta\,\right\}\,,
\end{equation}
as given by the Mean2 and Std2 values in Tab.\,\ref{TAB_CHAR3}. The
last statistic (\ref{MIN}) assumes that either a low $S_{\rm LEE}$ or
$S_\beta$ value already indicates substructure which seems to be
reasonable in the light of the discussion of individual clusters given
above.

\subsection{Comparison with Jones \& Forman classifications}\label{TEST2}

\begin{figure}
\vspace{0.0cm} \centerline{\hspace{-0.7cm}
\psfig{figure=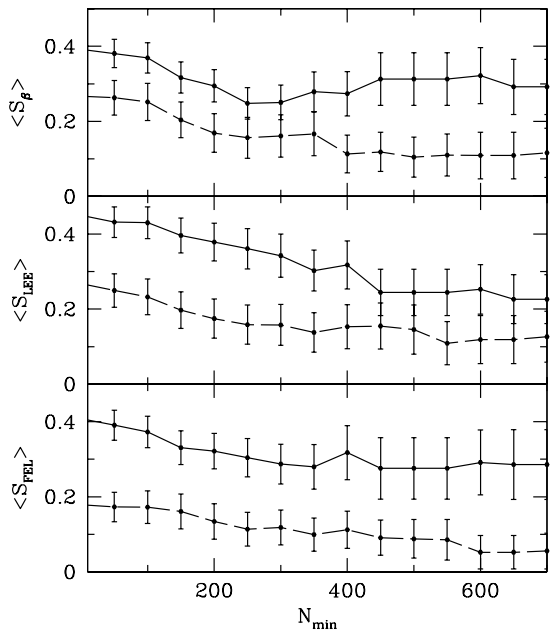,height=9.1cm}} 
\vspace{-0.50cm}
\caption{\small Average significances and their standard deviation as
a function of the number of RASS-3 photons of the 116 {\it Einstein}
clusters in common with the BCS and REFLEX cluster samples. The upper
curves in each of the three panels is obtained for the {\it Einstein}
clusters classified as single by Jones \& Forman (1999), the lower
curves for the {\it Einstein} clusters classified as double, primary
with small secondary, complex, elliptical, or offset center.}
\label{FIG_JONESFORMAN}
\end{figure}

Jones \& Forman (1999) used the iso-intensity contour plots of the
{\it Einstein} IPC X-ray emission of 208 targeted and serendipitously
found clusters with $z\le 0.15$ to classify their morphology into the
following categories. {\it Single}: no substructure or departures from
symmetry. {\it Double}: two subclusters of comparable size and
luminosity. {\it Primary with small secondary}: main subcluster at
least two times brighter than secondary. {\it Complex}: more than two
subclusters. {\it Elliptical}: elliptical X-ray surface brightness
contours. {\it Off center}: peak emission not in center defined by
lower surface brightness emission. {\it Galaxy}: emission dominated by
a galaxy. The {\it Einstein} and RASS-3 X-ray images have similar
angular resolutions (about 1 arc\-min). Moreover, the {\it Einstein}
classifications are restricted to clusters with redshifts $z\le
0.15$. Therefore, one should expect some correlation between the Jones
\& Forman and our classifications.

For the comparison of the two classification schemes we follow first
Jones \& Forman and decide that only the {\it Single} and {\it Galaxy}
types are regarded as regular and almost relaxed structures. The
remaining classes ({\it Double}, {\it Primary with small secondary},
{\it Complex}, {\it Elliptical}, {\it Off center}) are regarded as
clusters with significant substructure.  At this stage we do not doubt
the classification of Jones \& Forman because we only want to
illustrate the relation between two classification schemes.

In total 116 classified {\it Einstein} clusters are found in common
with the REFLEX and BCS sample and are regarded as a representative
test sample. Let us first show the relation of the three statistical
significances with the classifications of Jones \& Forman without
introducing an ad hoc threshold for the confidence
limit. Fig.\,\ref{FIG_JONESFORMAN} compares the average significances
obtained with each of the three statistics for the clusters classified
by Jones \& Forman as regular (upper continuous lines including their
formal $1\sigma$ Poisson errors) and substructured (lower dashed
curves). The average significances are computed for subsamples as a
function of the minimum number of X-ray photons, $N_{\rm min}$, in the
RASS-3 images.

\begin{figure}
\vspace{0.0cm} \centerline{\hspace{-0.7cm}
\psfig{figure=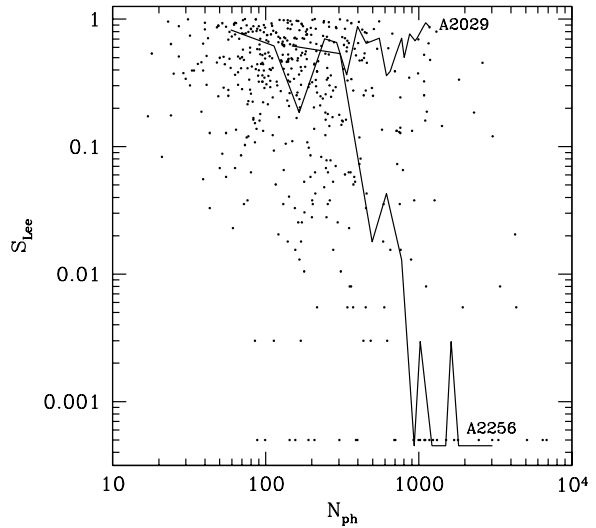,height=9.1cm}} 
\vspace{-1.50cm}
\caption{\small Distribution of LEE significances as a function of the
number of X-ray photons for REFELX$+$BCS clusters (points), for the
substructured cluster A2256 (lower continuous line), and for the
cooling flow cluster A2029 (upper continuous line). The lines
represent significances obtained with different numbers of X-ray
photons selected randomly from the X-ray images. Zero significance
values are set to $\sim 5\times 10^{-4}$.}
\label{FIG_DILUTE}
\end{figure}

For all $N_{\rm min}$ values a clear trend is seen that clusters
classified by Jones \& Forman as regular have on average larger
significances compared to the clusters classified by Jones \& Forman
as substructured. The largest differences between the mean $S$ values
of regular and substructured clusters are seen for the elongation
statistic. This indicates that elongation or ellipticity is
comparatively easy to classify both by eye and machine, especially for
large $N_{\rm min}$. Similar results supporting the robustness of this
criterion are given in Sect.\,\ref{TRUE}.

In the next step we want to illustrate the conditions under which our
substructure tests give a zero-order quantitative implementation of
the Jones \& Forman morphological classification scheme although the
X-ray data used for the two sets of classifications are
different. This also includes the introduction of a fixed confidence
level for substructure classification. Consistent with the results
obtained in Sect.\,\ref{TEST1} we now assume that clusters which are
classified by Jones \& Forman as elliptical are in fact
unstructured. In order to get higher signal-to-noise surface
brightness distributions and thus more reliable substructure
detections, we restrict the test sample to those clusters which have
at least 200 X-ray photons on the corresponding RASS-3 images. This
leaves 73 clusters for the final comparison.

From this test sample, 59 are classified as regular and 14 as
substructured by Jones \& Forman. We compare these numbers with the
results obtained with our substructure tests assuming that clusters
fulfilling the criterion min$\{S_{\rm LEE},S_\beta\}\le 0.05$ are
substructured.  In this case the substructure tests found 46 of the 59
clusters as regular and 11 of the 14 clusters as substructured. This
gives a fraction of 22 percent of the clusters with different
classifications. The rate is slightly improved to 21 percent when a
significance threshold of 0.01 is used.

Notice that the values of the SORs depend on the details of the sample
properties as well as on the actual substructure detection method so
that an exact coincidence of SORs obtained with different samples
cannot be expected. We conclude that the $\beta$ and LEE statistic are
found to provide useful tools for substructure analysis of RASS-3
X-ray images and can under reasonable conditions approximately verify
the Jones \& Forman classification scheme.

\begin{figure*}
\vspace{0.0cm} 
\centerline{\hspace{-9.0cm}
\psfig{figure=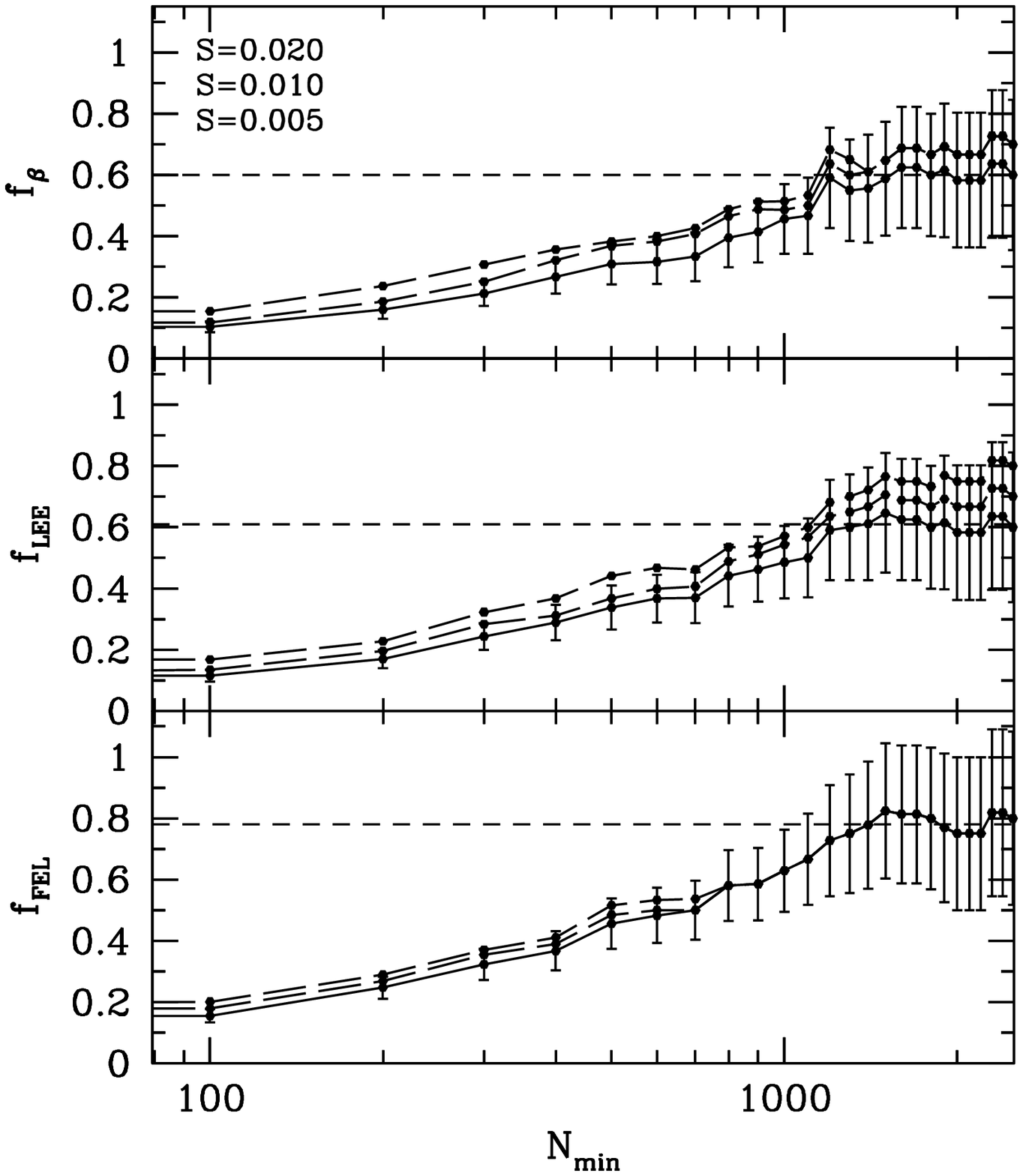,height=9.1cm}}
\vspace{-9.1cm} \centerline{\hspace{7.0cm}
\psfig{figure=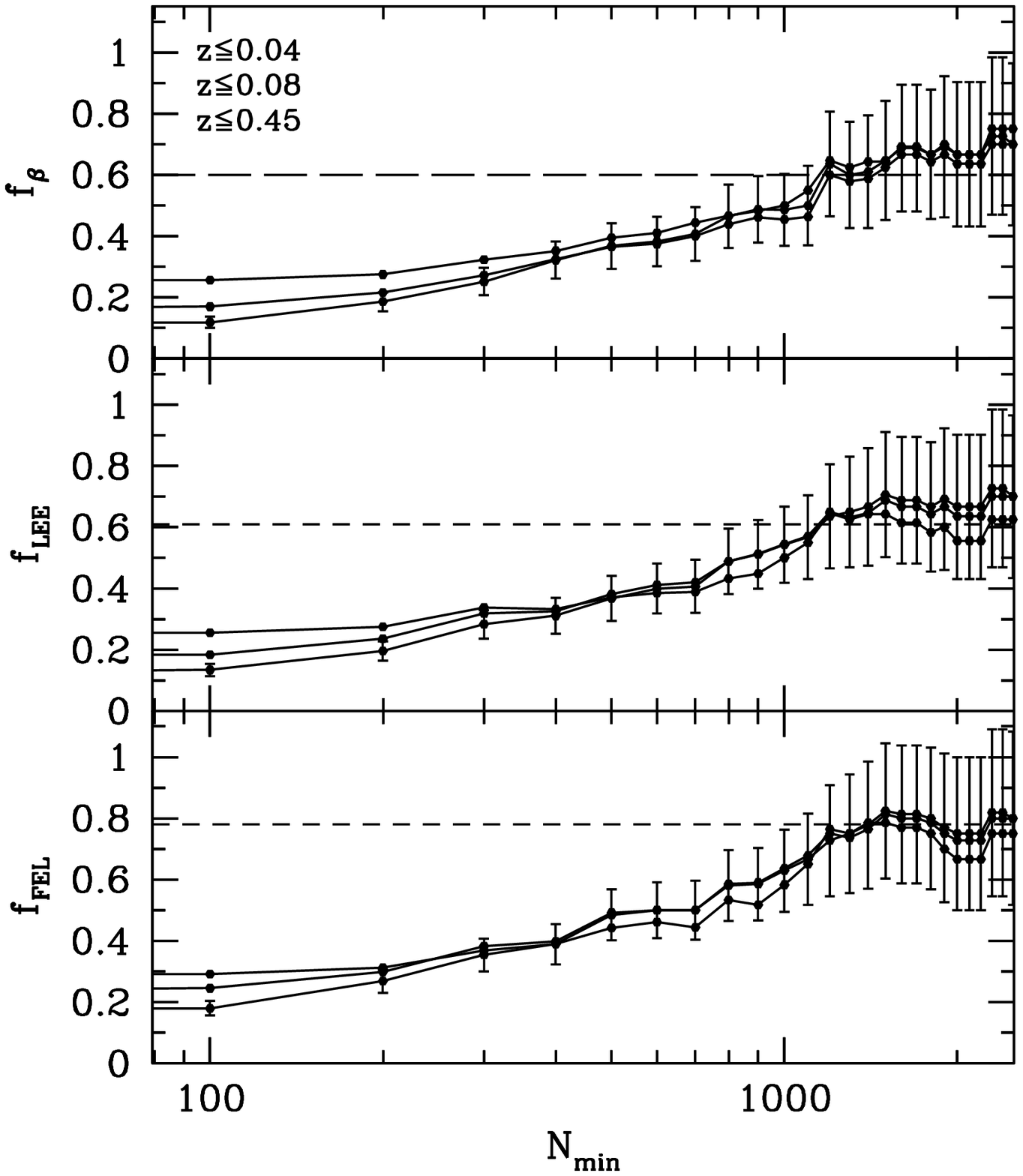,height=9.1cm}} 
\vspace{-0.50cm}
\caption{\small {\bf Left panel:} Fraction, $f$, of clusters as a
function of the minimum number of X-ray photons, $N_{\rm min}$, with
significances less or equal to the $S=0.005$ (lower curves with formal
Poisson error bars), 0.01 (middle curves) and 0.02 (upper curves)
level. In the upper panel the measurements for the $S=0.01$ and 0.02
levels degenerate above $N_{\rm min}=1\,400$ whereas for the lower
panel all curves degenerate above $N_{\rm min}=800$. The dashed
horizontal lines mark the plateau values obtained for clusters with
large numbers of X-ray photons. {\bf Right panel:} Same curves as left
panel for $S=0.01$ and the redshift limits $z=0.04$ (upper curves in
each panel, $\le 88$ clusters), $z=0.08$ ($\le 239$ clusters) and 0.45
($\le 470$ REFLEX$+$BCS clusters).}
\label{FIG_FREQ}
\end{figure*}

\section{Occurrence rates of clusters with substructure}\label{SELECT}

The observed or apparent frequency of clusters with significant X-ray
substructure is the basic quantity of our systematic study which can
be determined with high accuracy. The frequency distributions give
useful information about the role of substructure and merger events in
the nearby Universe, and provide several hints to important selection
effects which might bias substructure statistics in general.

For given spatial locations and size of main and subclusters, and for
given angular resolution of the X-ray instrument and aperture size,
the detectability of substructure depends on the number of X-ray
photons, $N_{\rm ph}$, used to trace the cluster surface brightness
distribution, and on the redshift of the
cluster. Fig.\,\ref{FIG_DILUTE} illustrates the $N_{\rm ph}$ effect,
plotting the statistical significances for the RE\-FLEX$+$BCS clusters
(dots) as a function of $N_{\rm ph}$. The superposed continuous lines
give sequences of statistical significances obtained for the Abell
clusters A2256 and A2029 for different numbers of photons selected
randomly from the original RASS-3 photon distribution. A2256 is a
well-known example of a cluster merger exhibiting a radio halo and
relic. Two subclumps are well-separated in the deeper RO\-SAT PSPC
pointing (Briel et al. 1991, supported by recent CHANDRA results given
in Sun et al. 2001), one maximum centered near the central cD
galaxy. The cluster is located close to the North Ecliptic Pole so the
correspondingly long ROSAT exposure time gives the comparatively large
number of 3\,000 X-ray photons within a circle of 1\,Mpc in RASS-3
(hard band, including background photons). A2029 is a cooling flow
cluster with a mass deposition rate $>550$ solar masses per year and a
cooling time $<3\,{\rm Gyr}$ (Peres et al. 1998). Buote \& Tsai (1996)
classified this cluster as regular and smooth and used it as a
prototype relaxed single cluster. Within a circle of 1\,Mpc about
1\,100 X-ray photons are found in RASS-3.

For other clusters, and the $\beta$ and FEL statistic, similar sample
paths as shown by the continuous lines in Fig.\,\ref{FIG_DILUTE} are
found. Therefore, if clusters are located in the upper left part of
the $S$-$N_{\rm ph}$ diagram with say $S>0.01$ corresponding to the 99
percent confidence level to reject the null hypothesis, this does not
neces\-sa\-rily mean that they do not have any substructure. It only
shows that under the actual detection conditions no statistically
significant statement about the tested substructure property can be
made. If more photons are collected the sample paths move to the right
of the $S$-$N_{\rm ph}$ diagram. But whether the path moves up or down
would depend on the actual merger situation, the size of the
subclusters, and on the cluster's redshift.

The observed occurrence rates, $f$, of the REFLEX$+$BCS clusters
obtained with the three substructure statistics are plotted in
Fig.\,\ref{FIG_FREQ} (left). For the morphological classifications the
significance thresholds $S\le 0.02$, 0.01, and 0.005 are
applied. Independent of the corresponding confidence limits the
observed fraction of substructured clusters increases with the minimum
number of X-ray photons, $N_{\rm min}$.  The comparison of the
corresponding diagrams for subsamples of REFLEX$+$BCS clusters
including an upper redshift limit (right panels of
Fig.\,\ref{FIG_FREQ}) shows that redshift-dependent effects become
important for minimum photon numbers $N_{\rm min}\le 500$.  They
increase when in addition to an upper redshift limit also a lower
limit is introduced. However, the fraction of high-$z$ clusters is
small so that their effect on the cumulative distributions is not
large.  Within the formal $1\sigma$ Poisson errors, the observed
fractions (complete sample) range between the lower limits of about
$10\pm 2$ percent and the plateau values of $60\pm 17$ percent for the
$\beta$ and LEE statistic, and $78\pm22$ percent for the FEL
statistic.

\begin{figure}
\vspace{0.0cm} 
\centerline{\hspace{-1.0cm}
\psfig{figure=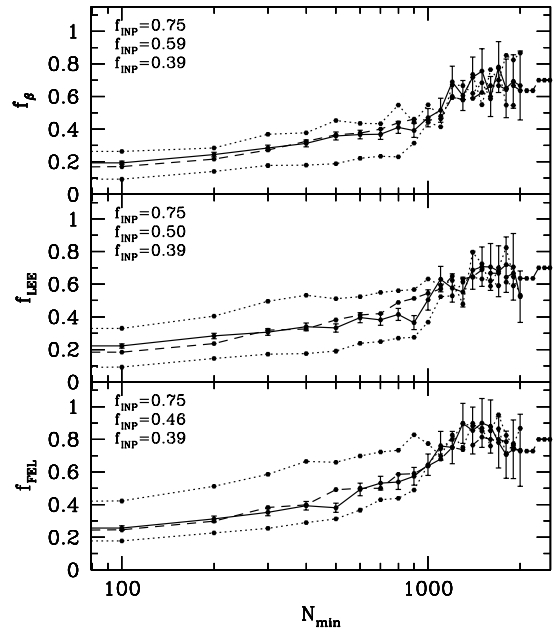,height=9.1cm}}
\vspace{-0.50cm}
\caption{\small Significance curves for REFLEX$+$BCS clusters with
redshifts $z\le 0.08$ (dashed lines) compared to the template samples
(continuous lines) covering the same redshift range. The input
substructure fractions of the control sample used to fit the
observations are shown in the upper left of each panel. The central
continuous line represents the best fit of the substructure fraction
to the observed curves. All curves are computed for the significance
threshold $S=0.01$.}
\label{FIG_FREQ_FITS}
\end{figure}

The amplitude and shape of the curves depend on the `true' SOR and the
X-ray flux and redshift distributions of the sample clusters which are
in the end determined by the structure formation process, on the X-ray
telescope and detector angular resolution, on the actual confidence
level used for the classification, etc.

\subsection{Towards unbiased substructure occurrence rates}\label{TRUE}

Less biased SORs of the REFLEX$+$BCS clusters can be determined with a
kind of template matching procedure where observed and model curves
similar to those plotted in Figs.\,\ref{FIG_FREQ} are compared and
their differences minimized. The model curves are computed with a set
of high signal-to-noise template clusters. This template set has by
construction a known substructure occurrence rate, $f_{\rm INP}$,
which can be changed by replacing substructured with regular templates
or vice versa. Partially dependent replicants of these templates with
different $N_{\rm ph}$ are generated by diluting their X-ray images
randomly in the same way as shown by the continuous sample paths in
Fig.\,\ref{FIG_DILUTE}.

For a proper comparison the enlarged set of template images generated
by dilution must have the same probability density, $P(N_{\rm ph})$,
as the REFLEX$+$BCS clusters. Precise comparisons would also equalize
the redshift distributions of observed and template clusters. This
requires, however, a large set of high signal-to-noise X-ray images
with at least 1000 X-ray photons at various redshifts. Such a template
set is not available yet so that we simply restrict the redshift {\it
range} to be the same for both observed and template clusters.

The matching procedure itself consists for each substructure criterion
of the following steps. In the first step the fractions $f(N_{\rm
min})$ are computed with the REFLEX$+$BCS sample (see
Fig.\,\ref{FIG_FREQ}, left) and for a fixed confidence limit. In the
second step an input substructure occurrence rate, $f_{\rm INP}$, of
the template sample is chosen. The substructure test is computed for
the diluted X-ray images. The images are randomly selected in a way
that observed and template sets have the same $P(N_{\rm ph})$
distributions. In the last step the curve, $f(N_{\rm min})$, is
determined and compared with the observed curve. Steps two and three
are iterated by changing $f_{\rm INP}$ of the templates until a useful
fit of the observed SOR curve by the template model curve is achieved.

Note that the application of a template sample based on RASS-3 data
guarantees that many technical selection effects which might affect
the REFLEX$+$BCS measurements are taken automatically into
account. The quality of the results depends on the size and
representativity of the template sample.

The results shown in Fig.\,\ref{FIG_FREQ_FITS} are computed for a
template sample of 37 clusters ($N_{\rm ph}\ge 1000$) with redshifts
$z\le 0.08$ and different input substructure fractions. Notice that
the classification of templates into substructured and regular
clusters is done by using literature information of the kind given in
Tabs.\,\ref{TAB_CHAR1} and \ref{TAB_CHAR2}, and by visual
inspection. As an example, curves are shown for a significance
threshold of $S=0.01$. The results obtained with the 239 RELFEX$+$BCS
clusters are limited to the same maximum redshift as the control
sample.  Each template cluster is diluted 20 times so that an
effective number of 740 partially dependent template clusters is used.

A simultaneous fit of the observed and model curves turns out to be
complicated because the shapes of the curves differ systematically
when specific $N_{\rm min}$ ranges are considered, indicating the
presence of some remaining systematic differences between observed and
template samples. One could, for example, expect that the template
sample might not be representative of the complete REFLEX$+$BCS
sample. Best (eye-ball) fits (continuous lines) give SORs ranging from
46 percent (FEL) to 59 percent ($\beta$).

The reference (dotted) lines give some information about the
sensitivity of the results to changes of $f_{\rm INP}$. They are
computed for $f_{\rm INP}=0.39$ and 0.75. The flatter FEL curves
suggest that elongation is less affected by $N_{\rm ph}$. This result
supports our evaluation of FEL for individual clusters
(Sect.\,\ref{TEST2}) and is quite important for our planned studies of
alignment effects of position angles of major cluster axes.

The final estimate of the `true' SOR is estimated by the formal mean
and standard deviation of the three input rates and gives 
\begin{equation}\label{SORC}
{\rm SOR}\,\,=\,\,52\pm\,7\,\,{\rm percent}\,.
\end{equation}
Note that this result is independent whether elongation is regarded as
substructure because FEL is only used as a link between template and
REFLEX$+$BCS clusters.

The nature of the merger events involved to yield the observed SOR is
not clear because the applied substructure tests do not give mass
estimates of the individual subclumps. Therefore, numerical
simulations are in preparation to decide statistically which merger
type (major merger or accretion) contributes at a given redshift and
flux to the observed SOR.

Another way to proceed further is the relative comparison of
subsamples derived from the REFLEX$+$BCS sample so that selection
effects introduced by the actual observing conditions partially cancel
out. Results obtained with this approach are described in the
following sections.

\section{Substructure density relation}\label{MORPH}

\begin{figure}
\vspace{-4.0cm} \centerline{\hspace{1.5cm}
\psfig{figure=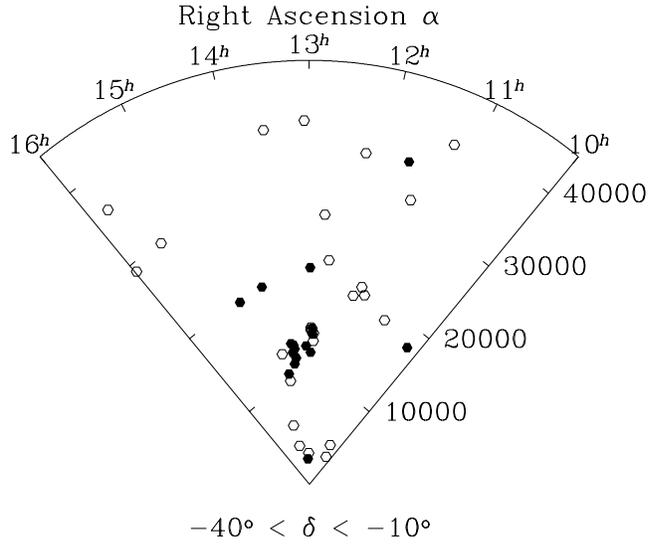,height=12.0cm}}
\vspace{-0.50cm}
\caption{\small Spatial distribution of clusters with ${\rm
min}\{S_\beta,S_{\rm LEE}\}\le 0.1$ (filled symbols) and $>0.1$ (open
symbols). The radial coordinate is in ${\rm km}\,{\rm s}^{-1}$. For
substructure detection a metric radius of 1\,Mpc is used.}
\label{FIG_CONE}
\end{figure}

Clusters in dense (supercluster) environments are expected to have a
higher probability to interact with neighbouring clusters or
filamentary structures connecting the cluster centers. If this
hypothesis is correct, larger fractions of clusters with distorted
X-ray surface brightness distributions and thus with subclusters are
expected in dense environments. 

As an illustrative example, Fig. \ref{FIG_CONE} shows the spatial
distribution of the clusters with ${\rm min}\{S_\beta,S_{\rm LEE}\}\le
0.1$ (filled symbols) and $>0.1$ (open symbols) in a region located at
the Shapley concentration. It is seen that clusters with a lower
significance to have a regular RASS-3 image (filled symbols) appear to
be preferentially more concentrated towards the core of the Shapley
supercluster. If the significance threshold of 0.1 is lowered to 0.01
basically all clusters expected to show substructure are located in
the Shapley concentration (only one cluster lies outside the
supercluster). The effect thus seems to be not very sensitive to the
actual value of the significance threshold applied. Similar but less
extreme trends are found for other supercluster regions.

\begin{figure*}
\vspace{0.0cm} 
\centerline{\hspace{-9.0cm}
\psfig{figure=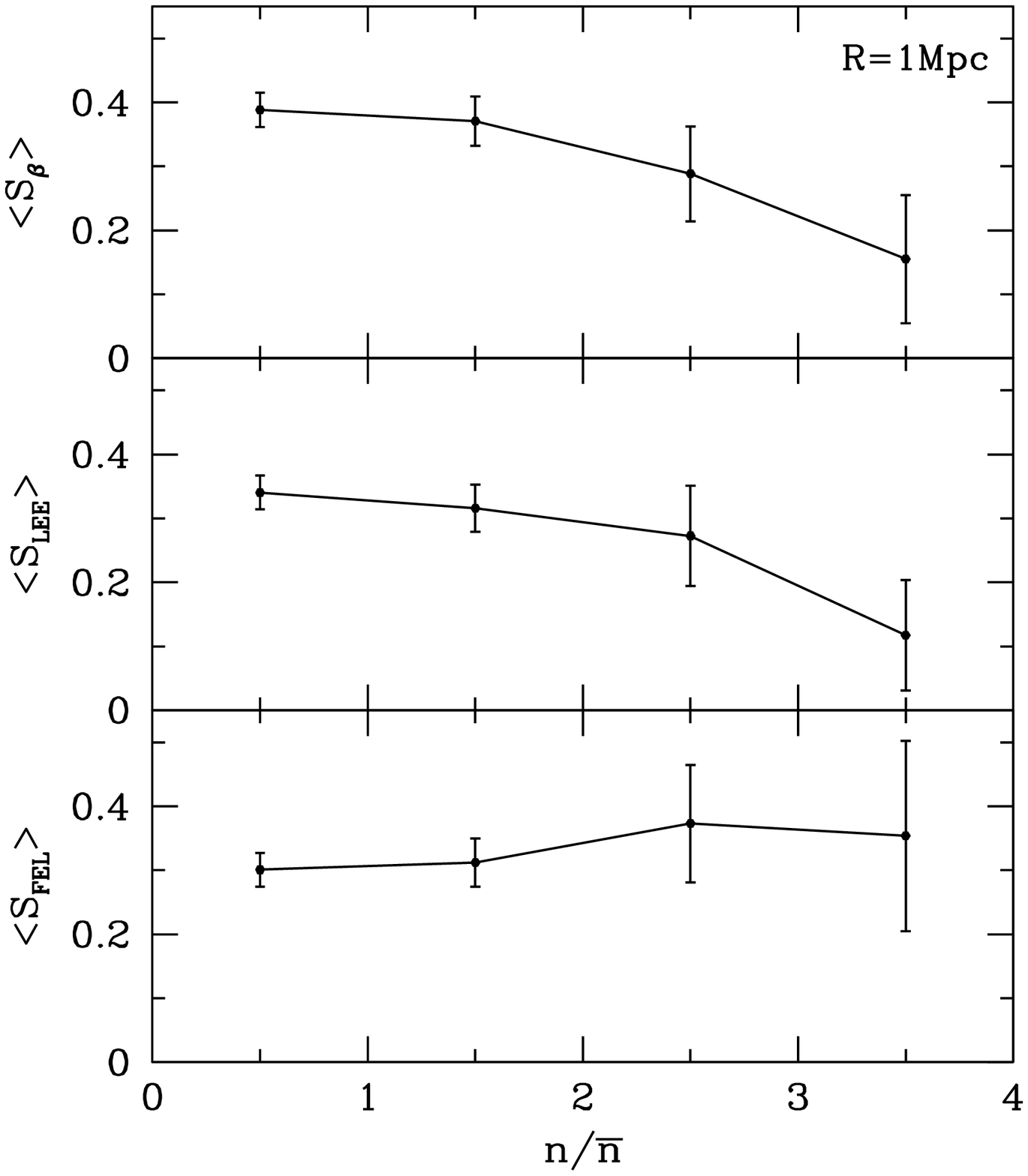,height=9.1cm}} 
\vspace{-9.1cm} \centerline{\hspace{7.0cm}
\psfig{figure=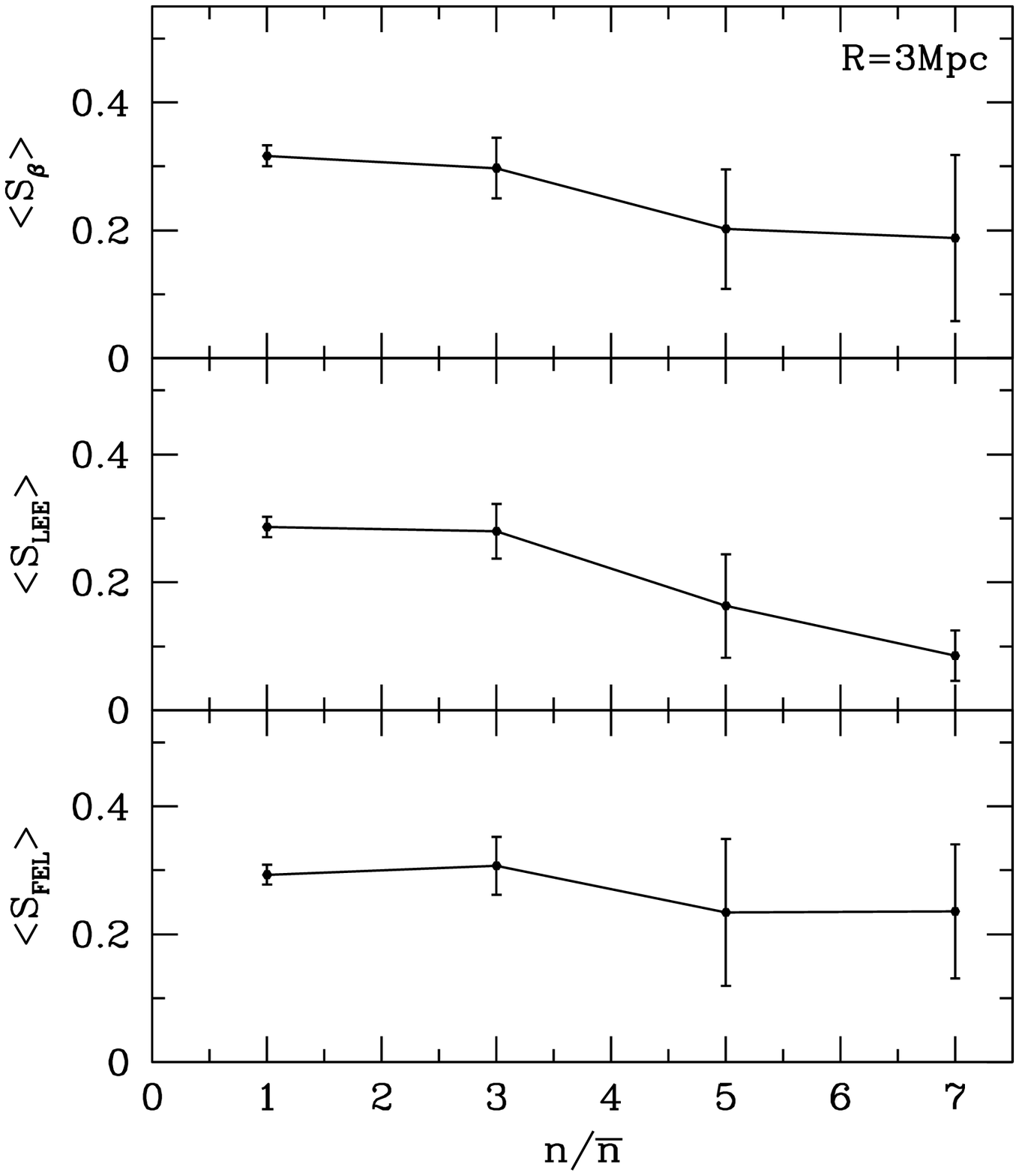,height=9.1cm}} 
\vspace{-0.50cm}
\caption{\small {\bf Left:} Average significances and their standard
deviation for the three statistical tests as a function of the local
cluster number density, normalized to the average density for an
aperture radius of 1\,Mpc. {\bf Right:} Same as left for an aperture
radius of 3\,Mpc.}
\label{FIG_MORPHDENS}
\end{figure*}

To quantify this effect and to test its statistical significance, mean
significances of $\beta$ and ${\rm LEE}$ (for completeness we also use
${\rm FEL}$) are computed for different local cluster number
densities. For volume-limited samples, the computation of local
cluster number densities would be straightforward. The REFLEX and BCS
cluster samples used for the present investigation are, however, X-ray
flux-limited and, although the samples are the largest yet available,
volume-limited subsamples derived from them have sample sizes smaller
than about 100 (see, e.g., Tab.\,1 in Schuecker et al. 2001). If we
would further subdivide these subsamples with respect to different
morphological classes the results would immediately suffer from
small-number statistics. Therefore, we decided to work with
flux-limited samples although the definition of redshift-independent
measures of local cluster number density becomes less well-defined.

As a measure of the local cluster number density around each cluster
we use $\bar{d}^{-3}$, where $\bar{d}$ is the mean of its 5 nearest
neighbour distances.  In order to correct for the redshift dependence
of the cluster number density in flux-limited cluster samples, we
normalize this number density by the average density obtained with the
same density estimator using all clusters in a redshift shell
$(z-\Delta z,z+\Delta z)$ centered on the cluster's $z$ value (see
comments given in Sect.\,\ref{DISCUSS}). The normalization has the
additional effect of compensating also for edge effects which are
known to distort next neighbour statistics (e.g., Cressie 1993).

The flux limit of REFLEX$+$BCS is quite bright and the segregation
effect quite large (see below) so that it is not necessary to correct
for local variations of the REFLEX and BCS survey sensitivity (e.g.,
RASS exposure convolved with galactic extinction). Note that cluster
X-ray fluxes are, however, corrected for galactic
extinction. Furthermore, no corrections for chance contamination of
cluster images by non-cluster sources are applied because results
obtained with different cluster subsamples drawn from the same parent
distribution are compared.

For the analysis of the RE\-FLEX$+$BCS sample the average cluster
number densities, $\bar{n}$, are computed within redshift shells with
$\Delta z=0.01$ centered at the cluster redshifts. The local density
contrasts are compared with the centers of optical superclusters given
in Zucca et al. (1993). Among others, the Perseus-Pisces,
Corona-Borealis, Hercules, Shapley, and Horologium superclusters are
clearly detected as regions with $n/\bar{n}>1$ for the majority of the
supercluster members.

Figure \ref{FIG_MORPHDENS} (left) shows the average substructure
significances as a function of the local cluster number density
contrasts excluding the very extreme contrasts where sample sizes are
small and the results quite noisy. The bars represent the formal
$1\sigma$ errors (excluding cosmic variance). It is seen that the
average significances of $\beta$ and LEE decrease with density
contrast. The effect is supported by subsamples of nearby clusters
with comparatively large numbers of X-ray photons. In addition, we
compare the sensitivity of the results to boundary effects possibly
introduced by the galactic plane. The same trends are found (although
with larger scatter) if only those clusters where the maximum distance
of the 5th neighbour cluster is far away from any survey boundary were
analyzed.

The same effect appears also on a larger metric scale when the
aperture radius for substructure detection is increased from 1\,Mpc to
3\,Mpc (see Fig.\,\ref{FIG_MORPHDENS}, right). Here we could follow
the trend to even higher density contrasts, although error bars get
quite large. For the latter case we can, however, not rule out that
neighbouring clusters not necessarily in the process of merging with
the programme cluster might artificially increase the SORs (notice
that the average significances are in general lower for the 3\,Mpc
apertures compared to the 1\,Mpc results). In addition, many clusters
have significance radii smaller than 3\,Mpc especially at lower
redshift so that the effect of background sources or local variations
of the X-ray background appears to be less clear.

To summarize, if the $\beta$ and LEE statistics are regarded as
reliable indicators for substructure (see Sect.\,\ref{FTESTS}) then
the results shown in Figs. \ref{FIG_CONE} to \ref{FIG_MORPHDENS}
clearly indicate that the fraction of substructured clusters increases
with local cluster number density.

Contrary to the results obtained with $\beta$ and LEE, the average
elongation significances are found to be almost insensitive to local
cluster number density contrasts. A decrease of $\langle S_{\rm
FEL}\rangle$ might be indicated on the 3\,Mpc scale, but the effect
seems to be not very strong. A constant value is just consistent with
the $1\sigma$ confidence limit. The different behaviour of FEL is not
yet fully understood. However, as noted in Sect.\,\ref{FTESTS}, the
presence of elongation is not necessarily restricted to merger events
so that low $<S_{\rm FEL}>$ values are expected in both high and
low-density regions, which could reduce a density-dependent effect.

\section{Substructures in halo, relic, and cooling flow clusters}\label{OBS1}

\begin{figure}
\vspace{0.0cm} \centerline{\hspace{-0.7cm}
\psfig{figure=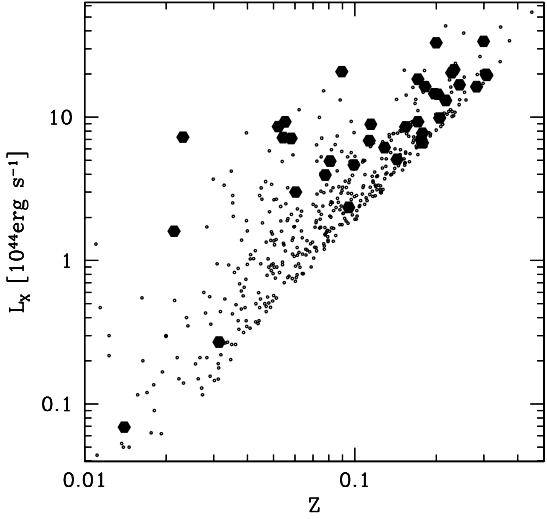,height=9.1cm}} 
\vspace{-1.50cm}
\caption{\small X-ray luminosity versus redshift for REFLEX$+$BCS
clusters (small points) and for halo and relic clusters (filled
circles).}
\label{FIG_LXZ}
\end{figure}

Radio halo and relic clusters are found among the most X-ray luminous
galaxy clusters. This is shown in Fig.\,\ref{FIG_LXZ} where the X-ray
luminosity in the energy band $0.1$-$2.4$\,keV is plotted as a
function of redshift for the RE\-FLEX$+$BCS reference sample (dots),
and the halo/relic clusters (filled circles). A threshold luminosity,
$L_{\rm X}=4.0\times 10^{44}\,{\rm erg}\,{\rm s}^{-1}$, might be
introduced above which most of the halo/relic clusters are located.
In the following we assume that the given halo/relic sample is
representative and compare the observed fractions of halo/relic
clusters with substructured RASS-3 X-ray images with the corresponding
results obtained for cooling flow and REFLEX$+$BCS clusters.

For a proper comparison of the frequencies of clusters with
substructure obtained with different samples one has to equalize the
efficiencies of substructure detection which might be different for
the samples.  Sect.\,\ref{SELECT} shows that the number of photons per
target, $N_{\rm ph}$, is the most crucial factor determining the
efficiency of substructure detection (the redshift bias will be
discussed below). The first step thus is to equalize the frequency
distributions of $N_{\rm ph}$ for the different samples. 
 
Both the halo/relic and the cooling flow samples have on average
higher X-ray fluxes yielding more X-ray photons per cluster target
compared to a typical RE\-FLEX$+$BCS cluster. Therefore, the
RE\-FLEX$+$BCS sample is in general less sensitive to substructure
detection compared to the other samples. The naive comparison of the
substructure tests would thus lead to a systematic underestimation of
the frequency of substructure for RE\-FLEX$+$BCS clusters.

In order to correct for the systematic effect one has to dilute a
larger reference sample so that its normalized cumulative probability
distribution function, $P(<N_{\rm ph})$, resembles the distribution
function of a smaller test sample. The dilution is performed as in
standard Monte Carlo experiments by randomly selecting from a uniform
distribution a number between 0 and 1, computing the corresponding
$N'_{\rm ph}$ value as given by the distribution function of the test
sample, and selecting the reference cluster with a $N_{\rm ph}$ value
next to $N'_{\rm ph}$. These steps are repeated until the replicant of
the reference sample has the same sample size as the test sample. In
order to get more representative results many replicants of the
reference sample are created, giving better estimates of the mean
values and their standard deviations.

\begin{figure}
\vspace{0.0cm} \centerline{\hspace{-0.7cm}
\psfig{figure=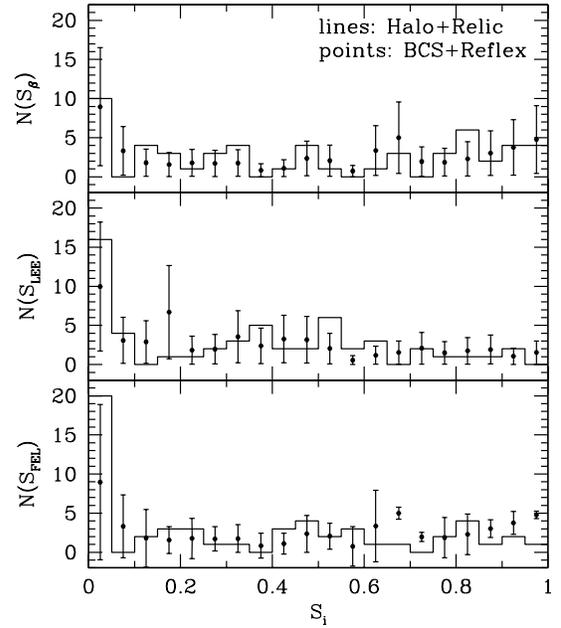,height=9.1cm}} 
\vspace{-0.05cm}
\caption{\small Frequency distributions of substructure significances
of halo/relic clusters (lines) and Reflex$+$BCS clusters (points).
The computation of the latter frequencies and their $1\sigma$ standard
deviations is described in the text.}
\label{FIG_HALO}
\end{figure}

\begin{figure}
\vspace{0.0cm} \centerline{\hspace{-0.7cm}
\psfig{figure=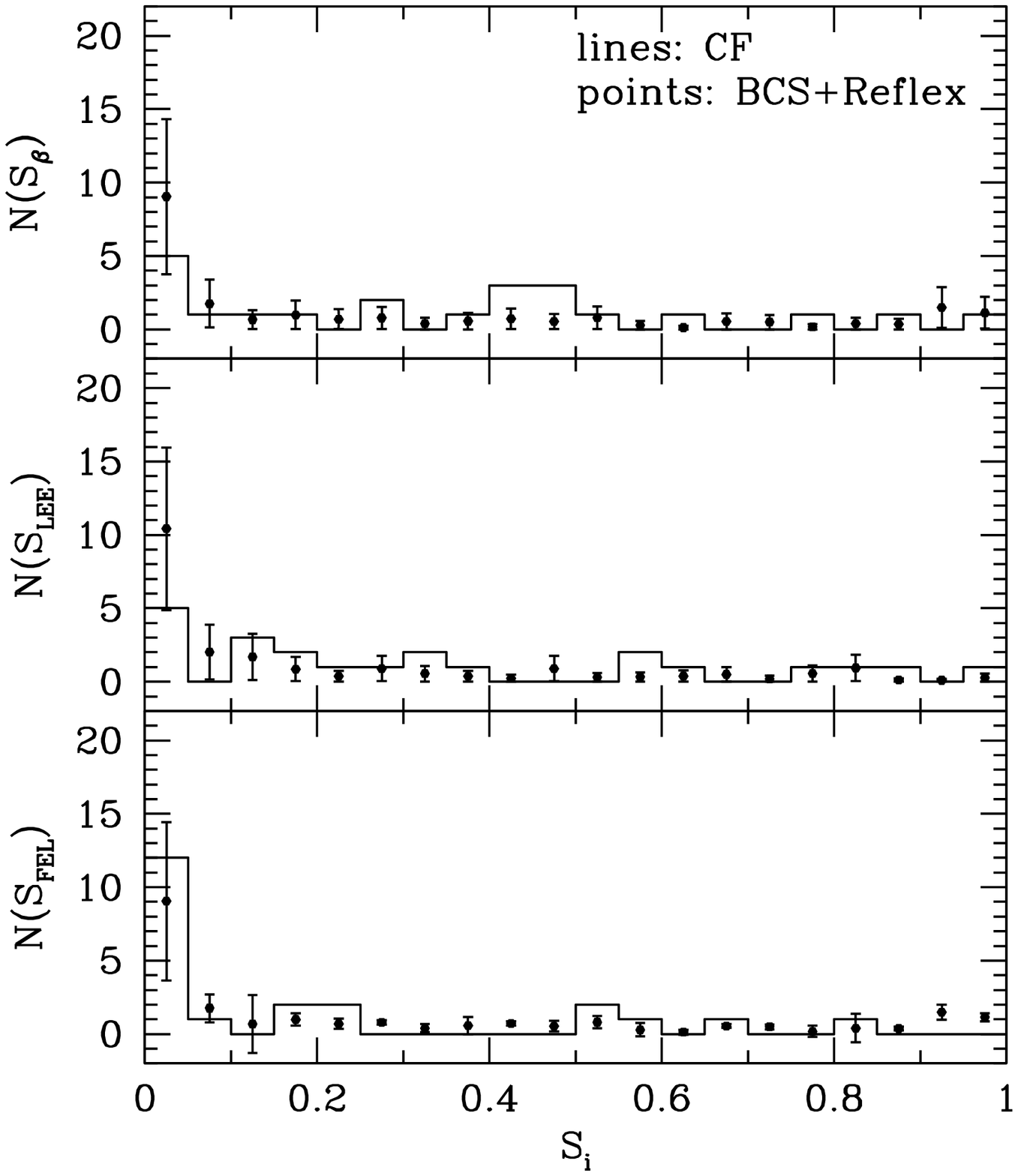,height=9.1cm}} 
\vspace{-0.50cm}
\caption{\small Frequency distributions of substructure significances
of cooling flow clusters (lines) and Reflex$+$BCS clusters (points).
The computation of the latter frequencies and their $1\sigma$ standard
deviations is described in the text.}
\label{FIG_COOL}
\end{figure}

\begin{figure}
\vspace{0.0cm} \centerline{\hspace{-0.7cm}
\psfig{figure=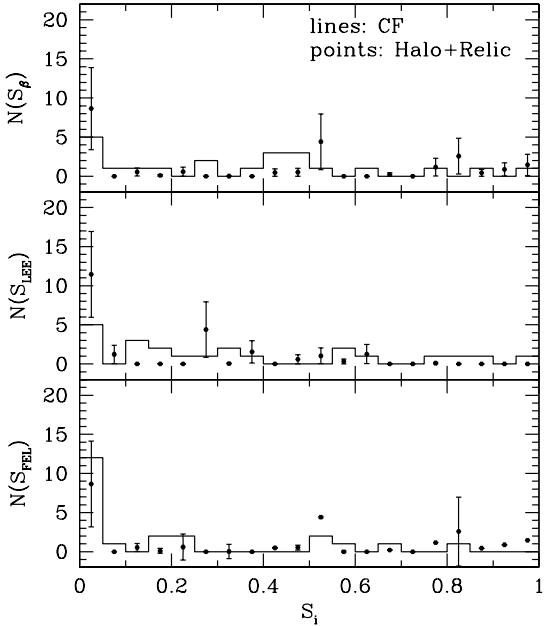,height=9.1cm}} 
\vspace{-0.50cm}
\caption{\small Frequency distributions of substructure significances
of cooling flow clusters (lines) and halo$+$relic clusters (points).
The computation of the latter frequencies and their $1\sigma$ standard
deviations is described in the text.}
\label{FIG_HALO_COOL}
\end{figure}

The resulting significance frequency distributions and their $1\sigma$
errors (points with error bars, see below) are given in
Figs.~\ref{FIG_HALO} to \ref{FIG_HALO_COOL}. Notice that the latter
figure compares halo/relic clusters with cooling flow clusters. In
this case the halo/relic sample is used as reference. The standard
deviations (error bars) are computed in the following way. As
mentioned above the reference sample is diluted so that its sample
size and photon number distribution resembles either the halo/relic,
or cooling flow sample. For each cluster type this is done 20\,000
times. The standard deviation is estimated from the scatter of the
numbers of clusters selected in each significance bin (a kind of
Bootstrap error).

It should be noted that the following analysis is restricted to a
discussion of the frequencies of the significances obtained for
circular symmetry ($\beta$), unimodality (LEE), and mirror symmetry
(FEL) for different cluster types. Therefore, the introduction of a
specific confidence level is not necessary. The results presented here
do also not depend on the specific binning used to count the
frequencies of the significance values. The following statements on
the significances are based on the frequencies and their errors
obtained in the first $S$ bin because here the differences in the
substructure behaviour are most interesting.  

The comparison indicates that halo/relic, cooling flow, and
REFLEX$+$BCS clusters do not strongly differ in their substructure
behaviour. The differences found in the RASS-3 images between these
types are all on the 1-$2\,\sigma$ level. However, the general picture
that cooling flow clusters appear to be more regular and halo/relic
clusters more often substructured is supported. In the following more
specific results are described.

Comparison of halo/relic with REFLEX$+$BCS clusters (see
Fig.\,\ref{FIG_HALO}): halo and relic clusters are found to have high
frequencies for all three substructure tests below $S=0.05$ compared
to REFLEX$+$BCS, suggesting that halo and relic clusters appear to be
on average more substructured compared to REFLEX$+$BCS
clusters. Whereas the excess is only marginal for $\beta$, about
$1\sigma$ excesses are found for both LEE and FEL. Halo/relic clusters
thus appear more often bi-modal and elongated compared to other
cluster types.

Comparison of cooling flow with REFLEX$+$BCS clusters (see
Fig.\,\ref{FIG_COOL}): cooling flow clusters are found to have lower
frequencies in the $S\le 0.05$ bin for $\beta$ and LEE whereas a small
excess is found for FEL compared to the REFLEX$+$BCS reference
sample. The deviations are on the $1\sigma$ level, but in the opposite
direction as found for the halo/relic clusters in
Fig.\,\ref{FIG_HALO}. The measurements thus suggest that clusters with
large cooling flow signatures appear to be more often circular
symmetric and unimodal. Significant elongations of their X-ray images
are found slightly more often compared to REFLEX$+$BCS clusters.

Comparison of cooling flow with halo/relic clusters (see
Fig.\,\ref{FIG_HALO_COOL}): cooling flow clusters are found to have
more often circular symmetric and unimodal RASS-3 images compared to
halo/relic clusters. Moreover, cooling flow clusters appear to be
marginally more often elongated.

\section{Discussion}\label{DISCUSS}

A systematic study of the morphologies of RASS-3 images of BCS and
REFLEX clusters is given. The two surveys provide the largest
presently available X-ray cluster samples and are expected to yield
statistically representative results. Our analysis shows that many
observational effects lead to systematic errors in the statistic of
merger events. After approximate corrections are applied an estimate
of the SOR of $52\pm 7$ percent is found. This number might also be
contaminated by chance superposition of point-like (background)
sources. However, the effects are expected to be small because a
comparatively small aperture radius of 1\,Mpc for substructure
measurements is used.

How does this SOR estimate compares with the results obtained with
similar projects in X-rays? As mentioned above, with 208 {\it
Einstein} IPC images Jones \& Forman (1999) find by visual inspection
a SOR of 41 percent. With 65 {\it Einstein} IPC images Mohr et
al. (1995) used the emission-weighted centroid variation for
substructure detection. Kolmogorov-Smirnov tests suggests that the
sample is representative. They found a SOR of 61 percent for the same
confidence level (99 percent) as used for the REFLEX$+$BCS sample.

It is thus seen that the three largest presently available systematic
X-ray cluster works give SORs of about 50 percent. However, the
conservative (formal) $3\sigma$ standard deviation of 30 percent
between the three estimates already indicates that there is still
considerable scatter between different samples and methods. The
conservative interval of substructure occurrence rates
\begin{equation}\label{SOR}
20\,\le \,f\,\le\,80\,\,{\rm percent}\,\,\,(99\%,{\rm
confidence\,\,range})\,,
\end{equation}
for nearby clusters with $z<0.15$ might give a realistic impression of
the current situation of statistical work on X-ray SORs.

The next step towards a physical understanding of the observed SOR
should be the determination of the mass scales of the subclumps and
the dynamical time scales involved. Note that the individual
contributions of major mergers and accretion to (\ref{SOR}) are not
given by the measurements. Obtaining quantitative estimates appears to
be quite difficult, even if the analysis would have been done with
better data and refined substructure tests. 

However, large sample sizes offer the possibility to calibrate the
substructure events at least in a statistical manner by the
application of the same substructure tests to both observed and
simulated X-ray cluster images distributed in flux and redshift in the
same way. The simulation will also include a realistic X-ray
background so that an artifical increase of the observed SOR through
background point sources (see above) is taken into
account. Simulations of this kind would establish the link between
substructure as defined by the various measures and the dynamical
state of a cluster (Schuecker et al., in preparation). Some
interesting statistical results deduced from the combination of
observational work and numerical experiments can be found in, e.g.,
Mohr et al. (1995).

Depending on the accuracy of this comparison one should also try to
investigate redshift-dependent effects where no information is
available yet. High-resolution N-body simulations of, e.g.,
Gottl\"ober, Klypin \& Kravtsov (2001) suggest an increase of major
merger rates by a factor of about 2 between redshift $z=0$ and 0.25.

Almost independent from the observational effects mentioned above the
relative comparison of cluster subsamples show that the fraction of
clusters with X-ray substructure turns out to be a function of local
cluster number density in the sense that the fraction of clusters with
low statistical significances for circular symmetric and unimodal
X-ray surface brightness distributions increases with local cluster
number density.  Such a substructure-density relation is expected when
the merging of subclusters causes the observed substructures. Although
the analyses are based on the largest presently available X-ray
cluster catalogue, we were forced to use flux-limited subsamples in
order to get statistically significant results. When larger X-ray
cluster samples are available, future studies will allow the
extraction of statistically significant {\it volume-limited}
subsamples. This will give a better defined cluster number density
with the additional benefit of rough mass estimates as obtained from
the mass X-ray luminosity relation of Reiprich \& B\"ohringer (2001).

The substructure-density relation of clusters appears to be analogous
to the morphology-density relation of galaxies. A related effect,
namely that dynamically young optical APM clusters are more clustered
than the overall cluster population was recently found by Plionis
(2001) thus supporting the present findings.

The relative abundances of radio halo, relic, and cooling flow
clusters with substructures could be studied in more detail.  Whereas
halo and relic clusters tend to show more often substructure in
RASS-3, cooling flow clusters show the opposite effect. Notice that
the comparisons of cooling flow clusters either with REFLEX$+$BCS
(Fig.\,\ref{FIG_COOL}) or with halo/relic clusters
(Fig.\,\ref{FIG_HALO_COOL}) give quite similar results supporting
their robustness because almost independent reference samples are
used. We thus regard the finding that cooling flow cluster show in
general more often regular and unimodal X-ray surface brightness
distributions than other cluster types as quite stable. Notice that a
fraction of $5/22=0.23$ cooling flow clusters show either $S_{\rm
LEE}=0$ or/and $S_\beta=0$ indicating that substructured cooling flow
clusters are to be expected. In addition, a fraction of $7/22=0.31$
have $S_{\rm FEL}\le 0.01$ indicating that one third of the cooling
flow clusters show significant elongation within the inner 1\,Mpc
radius, possibly impressed by the cD galaxy.

On the other hand, although halo and relic clusters show more often
bi-modal and elongated RASS-3 surface brightness distributions, they
appear to share more similarities with other cluster types than
cooling flow clusters.

Nevertheless, our analyses give additional support to the idea that
cluster mergers might trigger the formation of radio halos and relics
whereas pre-existing cooling flows might be disrupted by recent merger
events.

\begin{acknowledgements}
We thank Joachim Tr\"umper and the RO\-SAT team for providing the RASS
data fields and the EXSAS software, and the referee Jason Pinkney for
many useful comments. P.S. acknowledges the support by the
Verbundforschung under the grant No.\,50\,OR\,9708\,35.
\end{acknowledgements}

\end{document}